\documentclass[11pt]{article}

\usepackage[preprint]{acl}

\usepackage{times}
\usepackage{latexsym}

\usepackage[T1]{fontenc}

\usepackage[utf8]{inputenc}

\usepackage{microtype}

\usepackage{inconsolata}

\usepackage{graphicx}

\usepackage{graphicx}
\usepackage{xspace}

\usepackage[bb=pazo]{mathalpha}
\usepackage{subcaption}
\usepackage{wrapfig}

\usepackage{hyperref}       
\usepackage{url}            
\usepackage{booktabs}       
\usepackage{amsfonts}       
\usepackage{nicefrac}       
\usepackage{microtype}      
\usepackage{xcolor}         
\usepackage{amsmath}
\usepackage[T1]{fontenc}    
\usepackage{multirow}

\newcommand{\ours}{EffiR\xspace}

\title{Making Large Language Models Efficient Dense Retrievers}

\author{Yibin Lei\textsuperscript{1}\thanks{Equal contribution}, Shwai He\textsuperscript{2}$^*$, Ang Li\textsuperscript{2}, Andrew Yates\textsuperscript{3} \\
\textsuperscript{1}{University of Amsterdam} 
\quad \textsuperscript{2}{University of Maryland, College Park} \\
\textsuperscript{3}{Johns Hopkins University, HLTCOE}
\\
\texttt{y.lei@uva.nl}, 
\texttt{\{shwaihe, angliece\}@umd.edu}, \texttt{andrew.yates@jhu.edu}
}

\begin{document}
\maketitle
\begin{abstract}
    Recent work has shown that directly fine-tuning large language models (LLMs) for dense retrieval yields strong performance, but their substantial parameter counts make them computationally inefficient. While prior studies have revealed significant layer redundancy in LLMs for generative tasks, it remains unclear whether similar redundancy exists when these models are adapted for retrieval tasks, which require encoding entire sequences into fixed representations rather than generating tokens iteratively.
    To this end, we conduct a comprehensive analysis of layer redundancy in LLM-based dense retrievers. We find that, in contrast to generative settings, MLP layers are substantially more prunable, while attention layers remain critical for semantic aggregation.
    Building on this insight, we propose EffiR, a framework for developing efficient retrievers that performs large-scale MLP compression through a coarse-to-fine strategy (coarse-grained depth reduction followed by fine-grained width reduction), combined with retrieval-specific fine-tuning.
    Across diverse BEIR datasets and LLM backbones, EffiR achieves substantial reductions in model size and inference cost while 
    preserving the performance of full-size models.\footnote{Our code and models are available at \url{https://github.com/Yibin-Lei/EffiR}.}
\end{abstract}

\section{Introduction}

Dense retrieval models \citep{karpukhin2020dense, xiong2021approximate, hofst2021efficiently,izacard2022unsupervised, ma2024finetuningllama} map queries and documents into a shared dense vector space, enabling efficient similarity-based search. Compared to traditional sparse methods like BM25~\cite{robertson1995okapi}, dense retrievers offer stronger semantic matching capabilities and have shown superior performance across a variety of information retrieval benchmarks \citep{ msmarco,thakur2021beir, muennighoff2023mtebmassivetextembedding}.

\begin{figure}[t]
    \centering
    \includegraphics[width=\linewidth]{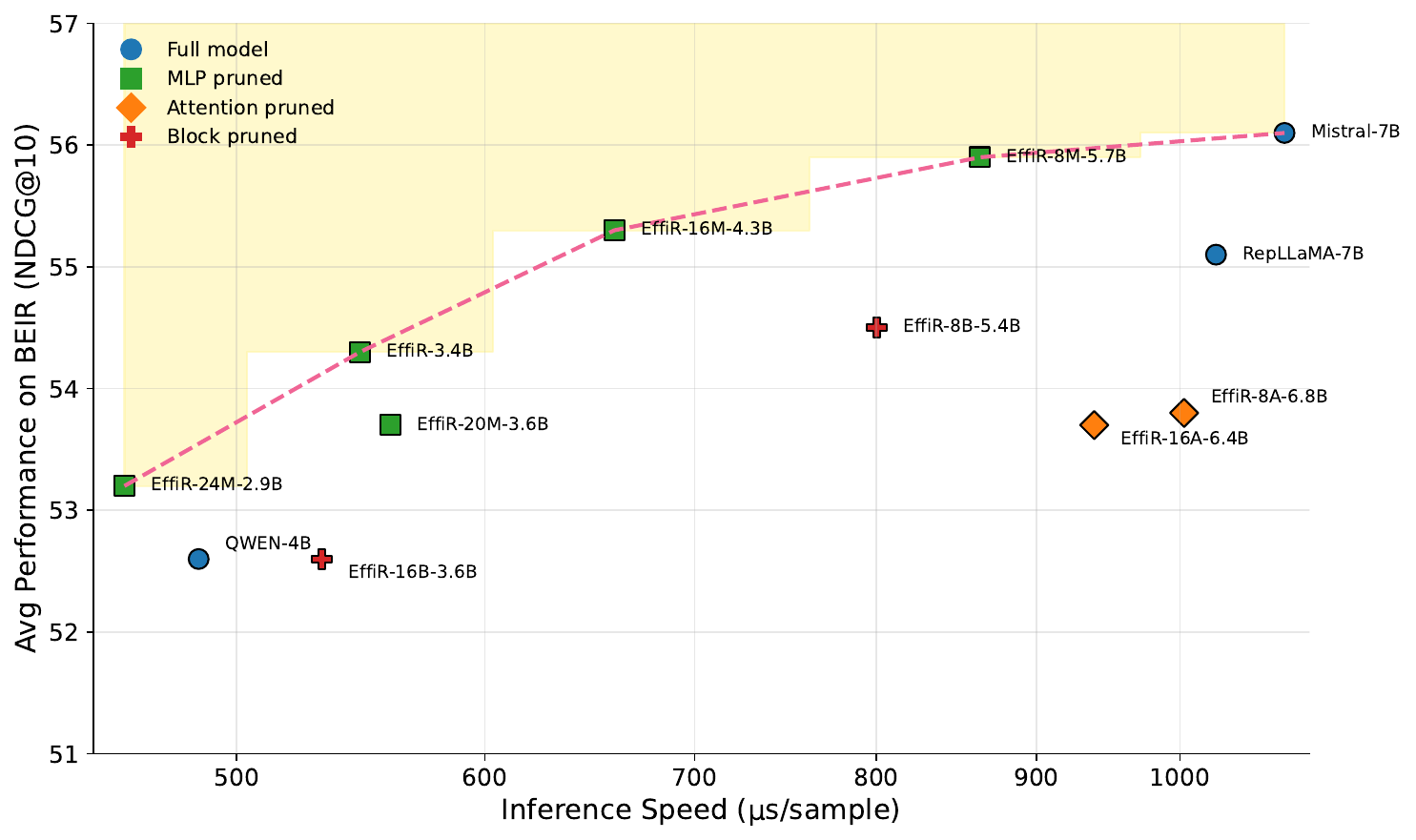}
    \caption{
    {Effectiveness–efficiency trade-off in LLM-based dense retrievers}. Each point shows a model’s BEIR performance vs. inference speed. All models are fine-tuned on MS MARCO. Marker types indicate compression strategies. EffiR builds efficient models based on Mistral-7B. For example, EffiR-20M-3.6B applies our EffiR method by dropping 20 MLP layers from Mistral-7B, then fine-tuning the remaining 3.6B parameters. MLP-pruned models (green squares) consistently lie near the Pareto frontier (dashed), showing strong efficiency with minimal accuracy loss.
    }
    \label{fig:figure-1}
    \vspace{-15pt}
\end{figure}

Large language models (LLMs) have recently emerged as powerful backbones for dense retrieval, producing high-quality text embeddings with strong generalization~\citep{ma2024finetuningllama}, multilingual capabilities~\citep{li2024makingtextembeddersfewshot}, and data efficiency~\citep{luo-etal-2024-large}. They also reduce reliance on large-scale retrieval-oriented pretraining~\citep{wang2024improving-text} and exhibit strong instruction-following abilities~\cite{sun-etal-2024-mair}. However, these benefits come with a significant computational cost: these models typically rely on large models with billions of parameters, making them impractical for real-time deployment.

Recent research shows that LLMs exhibit considerable layer redundancy in \textit{generative tasks}, where attention layers are prunable while MLP layers remain critical, enabling the removal of a substantial portion of layers with negligible degradation in performance~\citep{gromov2024unreasonableineffectivenessdeeperlayers, he2024matterstransformersattentionneeded, he2025understandingharnessingsparsityunified}. 
However, despite these findings, dense retrievers 
are still typically built by directly fine-tuning full LLM architectures without leveraging layer redundancy~\citep{wang2024improving-text, ma2024finetuningllama}, limiting their efficiency and practicality.
Moreover, dense retrievers serve a fundamentally different purpose from the pretraining objective of LLMs: instead of predicting the next token iteratively, they aim to produce a single, semantically meaningful representation for the entire input sequence. This distinction motivates examining: \textbf{(i) Do the architectures of foundation models also exhibit layer redundancy in retrieval tasks? (ii) If so, how does this redundancy differ from that of generative tasks? (iii) How can this redundancy be exploited to develop more efficient retrieval models?}

To investigate these questions, we conduct a systematic analysis of layer redundancy across multiple LLM backbones, considering two settings: directly pruning off-the-shelf retrievers and pruning followed by contrastive fine-tuning. 
We investigate how model performance changes when different layers are dropped, following layer-dropping techniques from prior work~\citep{he2024matterstransformersattentionneeded, he2025understandingharnessingsparsityunified}.
Interestingly, in contrast to findings in generative settings, we find that MLP layers, often viewed as repositories of factual knowledge~\citep{zhu2020modifyingmemoriestransformermodels,meng2022locating}, are more amenable to pruning in retrieval models. Conversely, attention layers exhibit less redundancy and cannot be removed as aggressively as in generative models~\citep{he2024matterstransformersattentionneeded, siddiqui2024adeeperlook}, as they 
play a crucial role in aggregating contextual information for fine-grained semantic matching.
Nonetheless, even with the higher redundancy of MLP layers, aggressive coarse-grained layer dropping alone leads to notable performance degradation beyond a certain compression ratio.

Motivated by the above findings, and given the substantial memory and computational overhead of MLP layers in retrieval scenarios, where auto-regressive decoding and key-value caching are not applicable, we propose \textbf{Effi}cient \textbf{R}etriever (\textbf{\ours}), a framework that performs large-scale retrieval-oriented MLP compression through a coarse-to-fine-grained strategy, followed by retrieval-specific fine-tuning. \ours employs two complementary compression stages: 
(i) \emph{coarse-grained depth reduction}, which removes entire low-importance MLP layers, and (ii) \emph{fine-grained width reduction}, which adaptively compresses the intermediate dimensions of the retained MLP layers.
As shown in Figure~\ref{fig:figure-1}, MLP-pruned models (e.g., EffiR-16M-4.3B) consistently lie on or near the Pareto frontier, demonstrating that substantial MLP compression can significantly improve efficiency with minimal performance degradation for retrieval. Notably, EffiR-3.4B, which applies our full coarse-to-fine framework, achieves higher BEIR performance than EffiR-20M-3.6B, which uses only coarse-grained layer dropping, despite having a smaller parameter count. This illustrates the advantage of combining coarse-grained depth reduction with fine-grained width compression, allowing us to retain key representational capacity while reducing redundancy more effectively than coarse methods alone. Further experiments demonstrate that our coarse-to-fine framework generalizes beyond Mistral and that EffiR is competitive with widely used pruning methods while pruning only MLP layers and providing substantial speedups through structural pruning.

\section{Related Works}
\paragraph{Dense Retrievers.} 
Early dense retrievers fine-tuned pre-trained language models (e.g., BERT~\citep{devlin-etal-2019-bert}) directly for retrieval tasks~\citep{karpukhin2020dense, izacard2022unsupervised, lei-etal-2023-unsupervised}, leading to a range of methods that incorporate advanced techniques like hard negative mining~\citep{xiong2021approximate, Wang2022TextEB, hofst2021efficiently}. More recently, LLMs have been adapted for dense retrieval~\citep{ma2024finetuningllama, weller2025promptriever, wang2024improving-text}, offering strong generalization, multilingual abilities, and less reliance on domain-specific supervision. However, these benefits come at the cost of significant computational overhead. 
Prior work has explored improving retrieval efficiency by reducing embedding dimensionality~\citep{kusupati2024matryoshkarepresentationlearning, lei-etal-2025-enhancing} or by accelerating similarity search using approximate nearest neighbor techniques~\citep{kumar2024ehi, sesmic}, but the encoding step, often the main computational bottleneck, has received comparatively less attention.
Beyond reducing embedding dimensionality alone (as in Matryoshka-style representations~\citep{kusupati2024matryoshkarepresentationlearning}), more recent 2D Matryoshka approaches train models that allow reducing both model layers and embedding dimensions, enabling further flexible effectiveness-efficiency trade-offs during encoding~\cite{starbucks,espresso}.
Recently, DRAMA~\citep{ma-etal-2025-drama} developed small dense retrievers from LLMs through a complex three-stage data augmentation pipeline (producing over 50 million synthetic training samples) combined with pruning across all model components, including both attention and MLP layers. In contrast, we discover that pruning MLP layers alone is sufficient and use this insight to construct a simple single-stage fine-tuning strategy using only standard MSMARCO data.

\paragraph{Redundancy in LLMs.} 
While increasing the depth of large language models significantly enhances their capacity~\citep{openai2024gpt4technicalreport, geminiteam2024gemini}, it also introduces considerable redundancy across layers. Recent studies have addressed this by identifying and removing less critical layers. For example, \citet{gromov2024unreasonableineffectivenessdeeperlayers} highlights the limited utility of deeper layers and advocates for dropping contiguous Transformer blocks. \citet{he2024matterstransformersattentionneeded} further examines the internal architecture of Transformer blocks and proposes fine-grained pruning strategies focused on attention layers, resulting in more effective layer reduction. Despite these advancements, existing work primarily focuses on generative models~\citep{jiang2023mistral7b, grattafiori2024llama3herdmodels} that perform token-level generation. In contrast, our work focuses on embedding models~\citep{chen-etal-2024-m3, Wang2022TextEB} that operate at the sequence level to produce fixed-length representations, and holistically investigates how redundancy manifests in this setting, offering new insights beyond the generative paradigm.

\begin{table*}[t]
\centering
\small
\vspace{-5pt}
\scalebox{0.8}{
\begin{tabular}{l|r|rr|rr|rr}
\toprule
 \textbf{\textit{E5-Mistral}} & Full-Model & Drop-8A & Drop-16A & Drop-8B & Drop-16B & Drop-8M & Drop-16M \\
 \#Params & 7.1B & 6.8B & 6.4B & 5.4B & 3.6B & 5.7B & 4.3B   \\
 \midrule
Arguana       & 61.6 & 55.3 & 0.1 & 39.0 & 9.3  & 59.0 & 8.4 \\
Climate-FEVER & 37.5 & 32.8 & 2.2 & 1.2  & 3.9  & 39.0 & 9.8  \\
DBPedia       & 48.6 & 44.8 & 0.6 & 19.0 & 9.7  & 44.7 & 15.2 \\
FEVER         & 88.3 & 83.8 & 9.5 & 36.6 & 22.5 & 89.6 & 56.6  \\
FiQA          & 57.2 & 51.8 & 2.4 & 9.1  & 8.1  & 53.5 & 14.0  \\
HotpotQA      & 75.6 & 67.5 & 2.8 & 43.0 & 9.6  & 75.5 & 18.0  \\
NFCorpus      & 38.7 & 36.1 & 2.6 & 11.4 & 17.5 & 37.4 & 18.2  \\
NQ            & 67.1 & 57.0 & 0.4 & 25.6 & 12.1 & 66.1 & 26.0  \\
Quora         & 89.3 & 87.1 & 0.6 & 21.5 & 26.8 & 89.1 & 66.0 \\
SCIDOCS       & 16.8 & 13.0 & 0.1 & 7.3  & 2.2  & 15.2 & 5.1 \\
SciFact       & 76.7 & 72.2 & 1.5 & 33.1 & 20.3 & 74.1 & 30.6 \\
TREC-COVID    & 86.6 & 85.2 & 12.6& 55.3 & 20.0 & 80.8 & 45.7  \\
Touche-2020   & 23.6 & 18.8 & 0.0 & 7.2  & 5.6  & 19.1 & 7.6  \\
\midrule
Average       & 59.1 & 54.3 & 2.7 & 23.8 & 12.9 & 57.2 & 24.7   \\
\bottomrule
\end{tabular}
}

\vspace{0.2cm}
\caption{Effectiveness (nDCG@10) and model sizes of E5-Mistral under different coarse-grained layer dropping strategies. ``Full-Model'' denotes the unpruned model. ``Drop-$k$A'', ``Drop-$k$B'', and ``Drop-$k$M'' indicate pruning $k$ self-attention layers, transformer blocks, and MLP layers using our coarse-grained layer dropping method, respectively. \textbf{No recovery training is applied.}}
\vspace{-0.6cm}
\label{tab:e5-mistral-prune}
\end{table*}

\section{How Retrieval Models are Different}

LLMs have shown promising performance in dense retrieval due to their ability to learn effective embedding representations. However, despite sharing the same underlying architectures, language modeling and dense retrieval serve distinct purposes. This section delineates these differences to motivate a distinct analysis of redundancy in dense retrievers.

\paragraph{Training Objectives.}

Dense retrievers and language models differ fundamentally in their training objectives. Language models are typically optimized with token-level predictive losses. For instance, causal language models adopt an autoregressive objective:
\begin{equation}
\mathcal{L}_{\text{LM}} = - \sum_{t=1}^{T} \log P(x_t \mid x_{<t}),
\end{equation}
where the model predicts each token $x_t$ conditioned on its leftward context $x_{<t}$. 
In contrast, dense retrievers are trained to produce semantically meaningful representations. A common approach is to optimize a \textit{contrastive loss} that brings matched pairs (e.g., a query and its relevant document) closer while pushing apart mismatched pairs. One widely used formulation is the InfoNCE loss:
\begin{equation}
\label{equ:infonce}
\small
\mathcal{L} = - \log \frac{\exp(\text{sim}(q, d^+)/\tau)}{\exp(\text{sim}(q, d^+)/\tau) + \sum_{j=1}^{N} \exp(\text{sim}(q, d_j^-)/\tau)},
\end{equation}
where $q$ and $d^+$ are the query and its corresponding positive document, $d_j^-$ are negative samples, $\tau$ is a temperature parameter, and $\text{sim}()$ typically denotes the similarity function between extracted representations, which is usually cosine similarity or inner product. These embeddings are usually derived from the model outputs using mean pooling or by extracting the hidden state of the last token.

\paragraph{Divergence in Inference Pattern.}

The fundamental difference leads to distinct inference patterns. Specifically, generative models operate autoregressively, where local information from previously generated tokens is often sufficient to predict the next token. In contrast, dense retrieval models process the entire input sequence once and require stronger global semantic aggregation to produce fixed-length representations. This divergence suggests that attention layers, which aggregate information across tokens, and MLP layers, which perform intra-token transformations, may contribute differently in the two modeling paradigms. 

On the other hand, unlike generative models, dense retrieval models obtain the representation in a single forward pass without the autoregressive decoding process, thereby eliminating the additional memory overhead (e.g., KV cache) associated with sequential attention computations. Moreover, MLP layers account for the majority of the parameters (e.g., 77.8\% in Mistral-7B~\cite{jiang2023mistral7b}), and are computationally intensive due to their operations in high-dimensional spaces. This suggests focusing more heavily on the MLP layers to achieve high compression rates and promote efficiency. 

\paragraph{Motivation.}

Despite the functional differences between generative and dense retrieval models, mainstream dense retrievers are still typically derived from general-purpose language models through post-finetuning, without modifying the underlying architecture. However, these architectures are originally pre-trained for generative tasks and may introduce unnecessary complexity when applied to retrieval-based applications.
This motivates us to examine whether general-purpose LLM architectures are overparameterized for retrieval and, if so, how their redundancy can be systematically leveraged to develop more efficient dense retrievers.

\section{{Are LLM-based Retrievers Redundant?}}
\label{sec:preliminary_redundant}

To examine redundancy in LLM-based dense retrievers, we apply the layer dropping method from~\citet{he2024matterstransformersattentionneeded} that focuses on removing the whole redundant layers to improve efficiency at scale. We study this by pruning three components of LLMs: self-attention layers, MLP layers, and full transformer blocks. While prior work on generative tasks~\citep{he2024matterstransformersattentionneeded, siddiqui2024adeeperlook} shows attention layers can often be pruned with little impact, it's unclear whether the same holds for embedding-based models like dense retrievers. 

\begin{table*}[t]
\centering
\small
\scalebox{0.63}{
\begin{tabular}{l|r|rr|rrr|rrrr|r}
\toprule
 \textbf{\textit{Mistral-7B}} & Full-Model & Drop-8A & Drop-16A & Drop-8B & Drop-16B & Drop-Last16B& Drop-8M & Drop-16M & Drop-20M & Drop-24M & Drop-16M8A \\
 \#Params & 7.1B & 6.8B & 6.4B & 5.4B & 3.6B & 3.6B & 5.7B & 4.3B & 3.6B & 2.9B & 4.0B   \\
 \midrule
Arguana       & 58.2 & 56.1 & 47.4 & 56.0 & 44.| & 43.0 & 56.9 & 54.5 & 51.4 & 45.6 & 53.2 \\
Climate-FEVER & 30.8 & 29.9 & 29.4 & 29.4 & 27.1 & 27.9 & 30.4 & 31.6 & 31.9 & 28.9 & 27.9 \\
DBPedia       & 44.2 & 40.1 & 42.1 & 40.2 & 41.5 & 42.0 & 43.9 & 42.2 & 37.4 & 41.3 & 39.2 \\
FEVER         & 83.5 & 79.1 & 77.4 & 81.9 & 76.3 & 76.5 & 82.6 & 83.3 & 79.7 & 80.6 & 79.9 \\
FiQA          & 45.9 & 45.5 & 43.5 & 44.8 & 41.8 & 43.9 & 45.8 & 44.9 & 43.1 & 40.3 & 43.7 \\
HotpotQA      & 70.5 & 68.1 & 66.2 & 67.9 & 65.1 & 65.4 & 71.0 & 70.1 & 67.9 & 67.5 & 67.9 \\
NFCorpus      & 34.7 & 31.6 & 34.2 & 32.5 & 35.3 & 37.8 & 33.3 & 35.4 & 33.1 & 36.3 & 33.9 \\
NQ            & 65.0 & 64.5 & 62.4 & 64.4 & 62.5 & 62.4 & 65.4 & 63.8 & 60.8 & 58.4 & 61.9 \\
Quora         & 83.1 & 83.3 & 87.9 & 84.2 & 88.2 & 86.8 & 85.9 & 83.7 & 86.5 & 87.0 & 84.3 \\
SCIDOCS       & 17.2 & 14.3 & 17.2 & 16.0 & 17.3 & 16.6 & 17.3 & 17.1 & 17.0 & 17.4 & 16.5 \\
SciFact       & 75.9 & 73.7 & 76.2 & 74.9 & 74.0 & 73.7 & 76.2 & 76.4 & 75.7 & 74.3 & 75.9 \\
TREC-COVID    & 84.1 & 83.0 & 86.3 & 82.8 & 82.2 & 85.6 & 83.7 & 84.2 & 85.7 & 83.3 & 85.3 \\
Touche-2020   & 35.7 & 30.7 & 28.4 & 33.0 & 28.3 & 27.1 & 34.4 & 32.3 & 28.1 & 31.1 & 28.7 \\
\midrule
Average       & 56.1 & 53.8 & 53.7 & 54.5 & 52.6 & 53.0 & 55.9 & 55.3 & 53.7 & 53.2 & 53.7  \\
\bottomrule
\end{tabular}
}
\vspace{0.2cm}
\caption{Effectiveness (nDCG@10) and model sizes of Mistral-7B variants \textbf{trained using MS MARCO} after coarse-grained layer dropping. We compare pruning of MLP layers (e.g., Drop-16M), attention layers (e.g., Drop-16A), full transformer blocks (e.g., Drop-16B), and their combinations (e.g., Drop-16M8A). Drop-Last16B denotes directly dropping the last 16 blocks.
\textbf{Results for LLaMA3-8B, Qwen-2.5-1.5B, Qwen-2.5-3B, Qwen-2.5-7B, and ModernBERT-base are provided in Appendices~\ref{app:llama3}.}}
\vspace{-10pt}
\label{tab:beir}
\end{table*}

\subsection{Redundancy Analysis via Layer Dropping}
Transformer-based models are stacked by multiple transformer blocks, each containing an attention layer and an MLP layer. So a $L$-layer model has $2L$ sub-layers and each sub-layer has a residual connection. 
For our analysis, we adopt a layer dropping strategy that removes sub-layers with minimal contribution to embedding quality. This requires estimating the importance of each sub-layer.
For the $l$-th sub-layer, we compute the importance scores: 
\begin{equation}
        S_{l} = M(x_l, x_{l+1}), \quad x_{l+1} = x_l + F_l(x_l),
\end{equation}
where $F_l$ denotes the $l$-th layer, and $M$ represents the matching metric, e.g., $l_2$-norm and cosine similarity. To assess the importance of the $l$-th layer, we compare the layer’s output $x_{l+1}$ with the input $x_l$. If the input closely matches the full output, it indicates that the incremental contribution of $F_l(x)$ is minimal, suggesting that the layer can be removed with limited impact on performance. 
Following prior work showing the effectiveness of cosine similarity for identifying redundant layers~\citep{gromov2024unreasonableineffectivenessdeeperlayers, he2024matterstransformersattentionneeded}, we adopt it as our importance metric, i.e., $M(x, y) = 1 - \text{Cosine}(x,y)$. 

Given the distinct roles of attention and MLP layers, where attention layers aggregate contextual information across tokens, while MLP layers perform intra-token transformations, we treat them as two separate groups during pruning. Specifically, we retain only the most important layers within each group. Let $S_{\text{Attn}}$ and $S_{\text{MLP}} $ denote the importance scores for the attention and MLP layers, respectively. The selected sets of layers are defined as:
\begin{equation}
    \small
    \mathcal{T}_{\text{Attn}} \leftarrow \text{Argmax}(\, S_{\text{Attn}} \,, k_\text{Attn}),
\end{equation}
\begin{equation}
    \small
    \mathcal{T}_{\text{MLP}} \leftarrow \text{Argmax}(\, S_{\text{MLP}} \,, k_\text{MLP}),
\end{equation}
where \( k_{\text{Attn}} \) and \( k_{\text{MLP}} \) denote the numbers of retained attention and MLP layers, respectively. The sets of selected layers are represented by $\mathcal{T}_\text{Attn}$ and $\mathcal{T}_\text{MLP}$, corresponding to the retained attention and MLP layers, respectively. 
The operator \( \text{Argmax} \) selects the top-\( k \) most important layers in each group. 

\subsection{Setup}
\label{sec:analysis_setup}
\paragraph{Pruning Setup.}
Following~\citet{he2024matterstransformersattentionneeded}, we compute importance scores for each layer using 256 validation samples from the C4 corpus~\citep{raffel2020exploring}, which is embedding and task-agnostic. We then prune the least important modules per category and evaluate the resulting models on 13 BEIR datasets~\citep{thakur2021beir} using nDCG@10. We also report parameter counts for each variant\footnote{In the paper, we exclude the language modeling head when reporting parameter counts, since text encoding does not require it.}. We consider two settings: (i) directly pruning off-the-shelf retrievers and (ii) pruning base models followed by contrastive fine-tuning.

\paragraph{Dense Retriever Training Setup.}
\label{sec:retrieval_train}
After compressing the base model, we fine-tune it for retrieval tasks. 
Following prior work~\citep{wang2024improving-text, ma2024finetuningllama}, we append a special <eos> token to each input text and use the hidden state of the final token as the text representation.
We train all dense retrievers using the InfoNCE loss defined in Equation~\ref{equ:infonce}, with the temperature hyperparameter set to 0.02, and additionally apply a distillation loss. All models are trained on MSMARCO data under identical settings to ensure fair comparison. Detailed training configurations are provided in Appendix~\ref{app:training_configutations}.

\subsection{Layer Redundancy Results}

\subsubsection{Pruning Off-the-Shelf Retrievers}
Table~\ref{tab:e5-mistral-prune} presents the results of directly pruning the off-the-shelf E5-Mistral model without any retraining. The results reveal several notable trends about redundancy in LLM-based dense retrievers. Unlike in generation tasks, where attention layers are typically more redundant, we find that pruning attention layers leads to a drastic collapse in performance, with dropping 16 attention layers nearly zeroing out retrieval scores across several datasets. This suggests that attention remains structurally vital for producing semantically rich embeddings. However, pruning MLP layers leads to a more graceful degradation: while performance drops are still noticeable, MLP-8 and MLP-16 retain moderate effectiveness on many datasets and yield significant parameter reductions. Interestingly, MLP pruning outperforms block-level pruning despite retaining less parameters (Drop-16M vs. Drop-8B), suggesting that MLPs offer a more efficient compression axis for dense retrievers.

\subsubsection{Pruning Followed by Fine-tuning}
To ensure a comprehensive examination, we also investigate an alternative setup: pruning the base model first, followed by contrastive fine-tuning. The training settings are detailed in Section~\ref{sec:retrieval_train}.

As shown in Table~\ref{tab:beir}, MLP layers exhibit the most redundancy. Dropping 16 MLP layers (\ours-16M) removes over 35\% of parameters with minimal performance loss (55.3 vs. 56.1). However, more aggressive pruning (\ours-20M, \ours-24M) leads to sharper degradation, highlighting the limitations of depth-only compression and motivating the width-aware, coarse-to-fine strategy used in \ours (Section~\ref{sec:slim}).

In contrast, pruning attention layers results in minimal parameter reduction but significant performance drops (e.g., \ours-16A drops to 53.7 nDCG@10). Block-level pruning (\ours-8B) falls between these extremes: less efficient than targeted MLP pruning and more stable than attention pruning, but lacking the precision of module-aware compression strategies.

These trends hold consistently across various models, including {LLaMA3-8B, Qwen-2.5-1.5B, Qwen-2.5-3B, Qwen-2.5-7B and ModernBERT-base} (see Appendices~\ref{app:llama3}), reinforcing our design principles for \ours: (i) prioritize MLPs as the primary compression dimension, and (ii) employ adaptive width-aware self-slimming rather than solely depth pruning to achieve superior efficiency-effectiveness trade-offs.

\section{EffiR: Efficient Retriever Training}
While results in Section~\ref{sec:preliminary_redundant} highlight the significant redundancy of MLP layers, they also indicate the limitations of depth-only pruning: removing too many layers degrades retrieval effectiveness substantially. To overcome this trade-off, we introduce EffiR (Efficient Retriever Training), a coarse-to-fine compression framework designed to  
explores redundancy from two complementary perspectives: \textbf{depth} and \textbf{width}. Depth reduction (i.e., layer dropping) removes the whole redundant layers to improve efficiency at scale, while width reduction adaptively compresses the remaining MLP layers to further enhance compactness while maintaining model performance.

\begin{table*}[t]
\centering
\small
\scalebox{0.7}{
\begin{tabular}{l|r|rrrr|rrrr|rr}
\toprule
  & RepLLaMA & Mistral-7B\textsuperscript{*} & Llama-1B\textsuperscript{*} & Gemma-2B\textsuperscript{*} & QWEN-4B\textsuperscript{*} & \ours-8A & \ours-16A & \ours-16M & \ours-20M & EffiR \\
 \#Params & 6.6B  & 7.1B     & 1.2B     & 2.6B      & 3.6B   & 6.8B & 6.4B & 4.3B & 3.6B & 3.4B   \\
  Query-Speedup & 1.05$\times$  & 1.00$\times$ & 6.71$\times$ & 3.24$\times$ & 2.22$\times$ & 1.08$\times$ & 1.15$\times$ & 1.64$\times$ & 1.93$\times$ & 1.97$\times$  \\
 Doc-Speedup & 0.98$\times$ & 1.00$\times$ & 6.91$\times$ & 3.59$\times$ & 2.18$\times$ & 1.08$\times$ & 1.17$\times$ & 1.55$\times$ & 1.80$\times$ & 1.82$\times$  \\
 \midrule
Arguana       & 48.6  & 58.2 & 49.7 & 57.5  & 48.6 & 56.1 & 47.4 & 54.5 & 51.4 & 52.8 \\
Climate-FEVER & 31.0  & 30.8 & 26.4 & 37.7  & 23.4 & 29.9 & 29.4 & 31.6 & 31.9  & 30.4 \\
DBPedia       & 43.7  & 44.2 & 38.2 & 37.7  & 43.1 & 40.1 & 42.1 & 42.2 & 37.4  & 43.0 \\
FEVER         & 83.4  & 83.5 & 81.8 & 78.3  & 80.2 & 79.1 & 77.4 & 83.3 & 79.7  & 82.0 \\
FiQA          & 45.8  & 45.9 & 37.8 & 40.6  & 40.1 & 45.5 & 43.5 & 44.9 & 43.1  & 42.8 \\
HotpotQA      & 68.5  & 70.5 & 65.2 & 65.7  & 65.0 & 68.1 & 66.2 & 70.1 & 67.9  & 67.6 \\
NFCorpus      & 37.8  & 34.7 & 32.3 & 33.7  & 35.6 & 31.6 & 34.2 & 35.4 & 33.1  & 35.9 \\
NQ            & 62.4  & 65.0 & 57.8 & 58.7  & 60.4 & 64.5 & 62.4 & 63.8 & 60.8  & 60.8 \\
Quora         & 86.8  & 83.1 & 82.4 & 77.7  & 83.3 & 87.9 & 84.6 & 83.7 & 86.5  & 83.3 \\
SCIDOCS       & 18.1  & 17.2 & 17.0 & 15.4  & 14.3 & 17.2 & 17.7 & 17.1 & 17.0  & 17.4 \\
SciFact       & 75.6  & 75.9 & 70.4 & 72.4  & 73.7 & 76.2 & 72.4 & 76.4 & 75.7  & 75.1 \\
TREC-COVID    & 84.7  & 84.1 & 81.7 & 76.9  & 83.0 & 86.3 & 84.0 & 84.2 & 85.7  & 83.3 \\
Touche-2020   & 30.5  & 35.7 & 29.3 & 28.3  & 29.1 & 32.3 & 30.7 & 28.4 & 28.1  & 31.7 \\
\midrule
Average       & 55.1  & 56.1 & 51.5 & 51.5 & 52.6 & 53.8 & 53.7 & 55.3 & 53.7 & 54.3 \\
\bottomrule
\end{tabular}
}
\vspace{0.2cm}
\caption{
Retrieval performance (nDCG@10), model size, and inference speedup of \ours and baselines on the BEIR benchmark. EffiR denotes the model trained using the full coarse-to-fine framework.
\ours-16A and \ours-16M denote variants of \ours that apply only coarse-grained pruning by dropping 16 attention and MLP layers, respectively, prior to training. \textsuperscript{*}Mistral-7B, Llama-1B, Gemma-2B, and QWEN-4B are trained with the same retrieval supervision with no compression applied.}
\vspace{-15pt}
\label{tab:effir}
\end{table*}

\subsection{Width Reduction via Self-Slimming}
\label{sec:slim}
In addition to the layer depth, the width of individual layers also contributes to the overall model size. In particular, the majority of parameters arise from MLP layers, which are formulated as: 
\begin{equation}
\small
\mathrm{MLP}(x) = W_{\mathrm{down}} \left( \mathrm{Act}(W_{\mathrm{gate}} x) \odot W_{\mathrm{up}} x \right) + x,
\end{equation}
where $W_{\mathrm{gate}} \in \mathbb{R}^{n \times d}$ and $W_{\mathrm{down}} \in \mathbb{R}^{d \times n}$ are the weight matrices of the MLP layer, and $\mathrm{Act}(\cdot)$ denotes an activation function. For simplicity, we omit the LayerNorm (\( \mathrm{LN} \)) in the formulation. 

The intermediate dimension $n$ is typically much larger than the hidden size $d$. For instance, Mistral-7B employs an intermediate size of 14{,}336 in its MLP layers, more than three times its hidden size of 4096.
While projecting the hidden states into higher-dimensional spaces enhances the model’s representational capacity, this architectural choice substantially increases the parameter count: MLP layers alone contribute to approximately 80\% of Mistral-7B’s total parameters.

To remedy this problem, we further propose self-slimming for width reduction across all MLP layers. Specifically, we propose an importance indicator trainable $\mathbf{z} \in \mathbb{R}^n$ at the intermediate neurons in an MLP layer. With this indicator, an MLP layer is reformulated: 
\begin{equation}
\small
\mathrm{MLP}(x) = W_{\mathrm{down}} \left( 
\mathrm{Relu}(\mathbf{z}) \cdot \mathrm{Act}(W_{\mathrm{gate}} x) \odot (W_{\mathrm{up}} x) \right) + x.
\end{equation}
Here, $\mathrm{ReLU}(\mathbf{z})$ serves two purposes. First, it ensures that the importance scores are non-negative, as negative values could lead to unintended cancellations. Second, it imposes a form of soft masking: a neuron with $\mathrm{ReLU}(z_i) \approx 0$ contributes minimally to the output and is a candidate for pruning. 
To ensure compatibility with the original MLP behavior, $\mathbf{z}$ is initialized as an all-ones vector, $\mathbb{1}^{n}$.

We then adopt a training-oriented approach to sparsify $\mathbf{z}$, making it trainable only during the self-slimming phase.  
Specifically, the overall training objective is defined as:
\begin{equation}
    \mathcal{L} = 
\mathcal{L}_\text{InfoNCE} + \lambda \mathcal{L}_\text{norm}.
\end{equation}
where $\mathcal{L}_{\text{norm}}$ is the  $\ell_0$-norm over $\mathrm{ReLU}(\mathbf{z})$, promoting sparsity in the active neurons,
and $\lambda$ is the corresponding regularization weight. 
The two loss terms jointly steer the scaling factors to  
(i) improve performance on the downstream retrieval task and  
(ii) remain highly sparse.  

However, since the $\ell_0$-norm is not differentiable and cannot be optimized directly via gradient-based methods, we use a sigmoid-based relaxation as a differentiable surrogate\footnote{Details on the relaxation are provided in Appendix~\ref{app:relaxation}.}. 

After only a few optimisation steps, the model learns a \emph{slim} activation pattern in which far fewer neurons are active. 
We then perform global pruning by ranking all scaling factors in 
$\mathrm{Relu}(\mathbf{z})$ across all MLP layers and freezing the least important values to 0 while setting the remaining ones to 1.
The resulting binary mask acts as a gating mechanism for neuron activation.
Next, we train this sparsified model with the $L_{\text{InfoNCE}}$ objective, under the normal dense retrieval training setting as described in Section~\ref{sec:retrieval_train}.
After training, we permanently remove all intermediate dimensions of MLPs whose corresponding scaling values are 0.

\section{\ours: Experimental Results}
In this section, we present comprehensive experiments demonstrating that \ours enhances the efficiency of LLMs for dense retrieval without compromising retrieval performance.

\subsection{Experimental Setup}

\paragraph{Model.}    
We develop \ours by applying our coarse-to-fine framework to the Mistral-7B-v0.1 model. 
In the first stage, we perform \textit{coarse-grained MLP layer dropping}, removing 16 MLP layers identified as least important, following the pruning setup described in Section~\ref{sec:analysis_setup}. In the second stage, we apply our fine-grained \textit{self-slimming} method to reduce the width of the remaining MLPs by 30\%. We also examine the generalizability of EffiR under different ratios and across alternative LLM backbones in Section~\ref{sec:ablation_self_slimming}.

\vspace{-0.2cm}
\paragraph{Evaluation.}
In line with the evaluation in Section~\ref{sec:preliminary_redundant}, we evaluate all models on the BEIR benchmark, using nDCG@10 as the metric.
To assess efficiency, we measure both the total parameter count and \textbf{inference speedup relative to the full Mistral-7B model}. Inference speedup is measured separately for query and document encoding using 1,000 randomly sampled inputs from the NQ dataset, executed on a single H100 GPU using the HuggingFace Transformers library, with \texttt{torch.compile} applied to the models.

\vspace{-0.2cm}
\paragraph{Baselines.}
We compare \ours against RepLLAMA~\citep{ma2024finetuningllama}, a strong LLM-based dense retriever trained on the MS MARCO dataset~\citep{msmarco}, the same dataset used to train our models. 
To isolate the effectiveness of our training framework, we also evaluate a set of small LLMs with similar release periods to Mistral-7B, trained under identical setups. These include LLaMA-3.2-1B~\citep{grattafiori2024llama3herdmodels}, Gemma-2-2B~\citep{gemmateam2024gemma2improvingopen}, and Qwen-1.5-4B~\cite{bai2023qwen1.5technicalreport} models. All baselines are trained using the identical configurations as described in Section~\ref{sec:retrieval_train}. In addition, we report results for layer dropping variants in which only coarse-grained layer dropping is applied prior to training.

\subsection{Main Results}

As shown in Table~\ref{tab:effir}, {\ours} achieves strong retrieval performance while significantly reducing model size and inference cost. With an average nDCG@10 of 54.3 across the BEIR benchmark, \ours closely matches the original Mistral-7B model (56.1), despite using only \textasciitilde48\% of its parameters and achieving a 1.97$\times$ query-side speedup.
\ours also consistently outperforms similarly sized small LLMs such as LLaMA-1B, Gemma-2B, and Qwen-4B, though trained under the same retrieval supervision. This highlights the strength of our framework: rather than relying on compact pretrained small models with limited capacity, \ours starts from a larger model and applies retriever-aware compression that preserves capacity in critical components while eliminating redundant computation.

In addition to outperforming general small models, EffiR improves over intermediate coarse-grained pruned variants like \ours-16M and \ours-20M, suggesting that our combined approach of coarse-grained layer dropping followed by fine-grained slimming is more effective than only using the coarse-grained method. This confirms the importance of designing compression methods tailored to the retriever's architectural priorities. 
\begin{figure*}[h]
  \centering
  \begin{subfigure}[t]{0.42\linewidth}
    \centering
    \includegraphics[width=\linewidth]{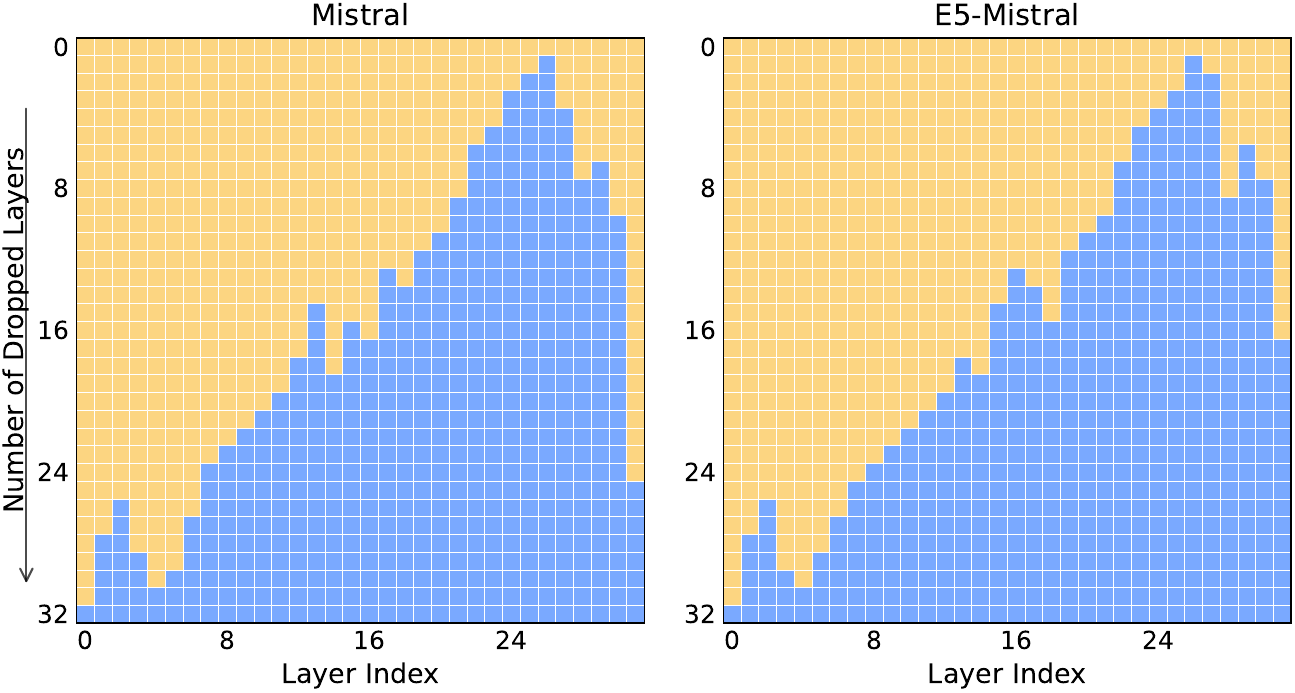}
    \caption{MLP Layer Dropping Heat Maps}
    \label{fig:drop_order_block}
  \end{subfigure}
  \hfill
  \begin{subfigure}[t]{0.42\linewidth}
    \centering
    \includegraphics[width=\linewidth]{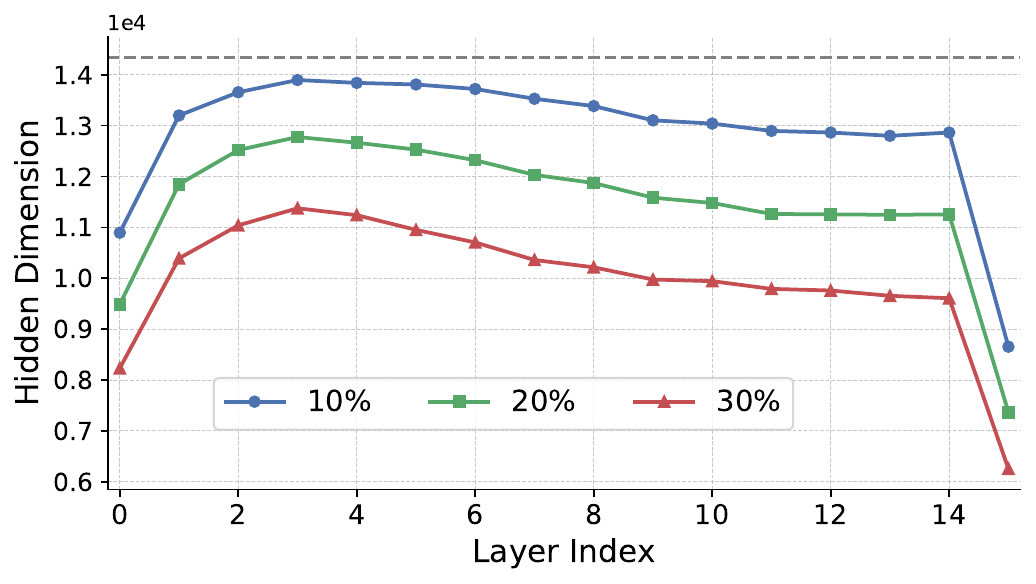}
    \caption{Layer-wise hidden dimension counts after width reduction under different reduction ratios.}
    \label{fig:drop_order_mlp}
  \end{subfigure}
  \hfill

  \caption{Analyzing compression behavior across layers.}
  \label{fig:drop_order}
  \vspace{-0.5cm}
\end{figure*}

\section{Analysis}

\subsection{Width Reduction vs. Layer Dropping}
\label{sec:ablation_self_slimming}
\begin{figure}[h]
    \centering
    \includegraphics[width=\linewidth]{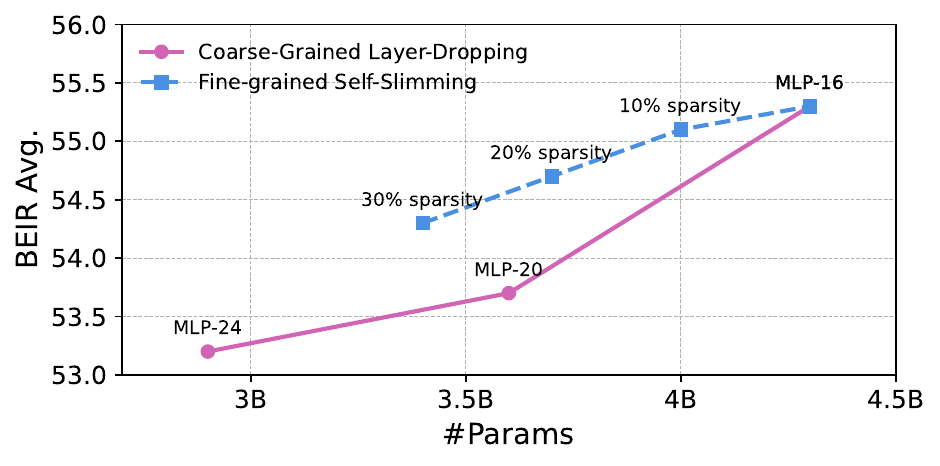}
    \caption{
    Comparison of coarse-grained layer dropping and fine-grained self-slimming, both starting from the same 16-layer dropped base Mistral-7B model. 
    }
        \vspace{-5pt}
    \label{fig:self-slimming}
\end{figure}

To analyze the impact of different compression stages, we compare coarse-grained MLP layer dropping with the fine-grained self-slimming method (Figure~\ref{fig:self-slimming}), starting from the same base: dropping 16 MLP layers from the Mistral model, a strong initial trade-off.

\begin{figure}[h]
    \centering
    \includegraphics[width=\linewidth]{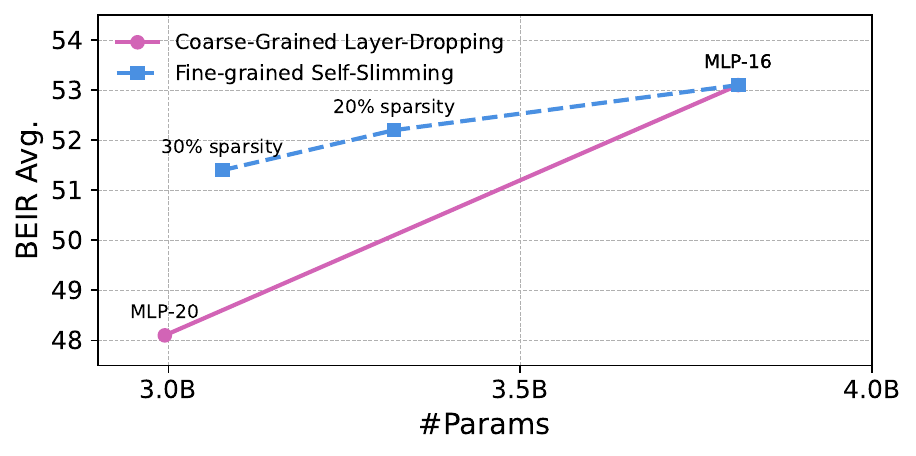}
    \caption{
    Comparison of coarse-grained layer dropping and fine-grained self-slimming, both starting from the same 16-layer dropped base Qwen2.5-7B model. 
    }
    \label{fig:qwen_slim}
    \vspace{-5pt}
\end{figure}

From this point, we explore two paths: (1) dropping additional layers (\ours-20M, \ours-24M), and (2) applying self-slimming to progressively reduce MLP width by various sparsity ratios. The trade-off is clear: further depth pruning reduces size but degrades performance, while self-slimming achieves a better efficiency-effectiveness balance. For example, 30\% self-slimming outperforms \ours-20M with fewer parameters. 
Results on Qwen2.5-7B (Figure~\ref{fig:qwen_slim}) exhibit the same overall pattern.
These results support the two-stage design of \ours: use coarse-grained pruning to reduce redundant depth, then apply fine-grained width reduction for precise, quality-preserving compression.

\subsection{Which Layers are More Redundant?}
\paragraph{Layer Dropping.} We visualize the layer-wise redundancy of MLP using dropping-order heat maps for both Mistral-7B and its retrieval-tuned variant E5-Mistral, as shown in Figure~\ref{fig:drop_order_block}. Each cell indicates whether a layer is dropped (blue) or retained (orange) during top-$k$ pruning. Notably, the redundancy patterns remain largely consistent after retrieval fine-tuning: the E5-Mistral variant shows similar trends to the base Mistral-7B model. Later layers are generally more prunable than earlier ones, highlighting a degree of overparameterization toward the top of the model.

\paragraph{Self-Slimming Width-Reduction.}

Figure~\ref{fig:drop_order_mlp}  presents the results of layer-wise hidden dimension counts under different overall sparsity ratios. We observe a consistent trend: deeper layers tend to be sparser than shallower layers among the remaining layers. This aligns with our layer-dropping results, where deeper layers are less important and are pruned earlier.

\subsection{Comparison with Pruning Methods}

\begin{table}[h]
\small
\centering
\setlength{\tabcolsep}{3.5pt}
\setlength{\abovecaptionskip}{0.2cm}
\setlength{\belowcaptionskip}{-0.0cm}
\begin{tabular}{@{} l c c c @{}}
\toprule
Method & Sparsity & Order & BEIR Avg.\\
\midrule
\multirow{2}{*}{Wanda} 
  & \multirow{2}{*}{50\%} 
  & Finetune-then-Prune & 49.8 \\
  &  & Prune-then-Finetune & 54.1 \\
\midrule
\multirow{2}{*}{SparseGPT} 
  & \multirow{2}{*}{50\%} 
  & Finetune-then-Prune & 50.5 \\
  &  & Prune-then-Finetune & 54.7 \\
\midrule
EffiR 
  & 52\% 
  & Prune-then-Finetune & 54.3 \\
\bottomrule
\end{tabular}
\caption{Comparison with sparsification-based pruning methods. Note the sparsity of Wanda and SparseGPT under prune-then-finetune is slightly higher than 50\% since the LoRA introduces additional parameters.}
\label{tab:wanda}
\vspace{-5pt}
\end{table}

We compare \ours with Wanda~\cite{sun2024awanda} and SparseGPT~\cite{sparsegpt}, two widely used pruning methods that induce parameter sparsity in both attention and MLP layers.
To ensure a comprehensive comparison, we evaluate them under both finetune-then-prune and prune-then-finetune settings, using the same data as \ours. Specifically, we use their 2:4 structured pruning veriants, which prunes at the block level and can leverage optimized sparse matrix operations to achieve real acceleration, achieving $1.24\times$ speedup on LLaMA-7B as reported. 

As shown in Table~\ref{tab:wanda}, under prune-then-finetune, both Wanda and SparseGPT achieve performance comparable to EffiR. However, their speedups rely on sparse-matrix acceleration rather than structural model changes, and the runtime improvement is not linear with sparsity (e.g., 50\% sparsity yields only a 1.24$\times$ speedup for LLaMA-7B, as reported in the Wanda paper).  
In contrast, EffiR performs hard pruning that removes entire layers and reduces hidden dimensions, which produces real reductions in computation and achieves substantially higher speedup.
Moreover, sparse-matrix acceleration requires specialized kernels and hardware support and still loads the full parameter set, offering no memory savings. EffiR avoids these limitations, making it more practical and deployment-friendly.

\begin{table}[h]
\small
\centering
\setlength{\tabcolsep}{3.5pt}
\setlength{\abovecaptionskip}{0.2cm}
\setlength{\belowcaptionskip}{-0.0cm}
\begin{tabular}{@{} l cc @{}}
\toprule
Method & \#Parameters & BEIR Avg.\\
\midrule
LlaMA2-7B & 6.6B & 55.7 \\
Sheared-LLaMA2&2.6B&49.6 \\
\midrule
EffiR-20MLP&3.9B&53.0 \\
\quad \textit{w/ $-30\%$} &3.4B&52.5 \\
\quad \textit{w/ $-50\%$}&3.1B&49.8 \\
EffiR-24MLP&3.4B&49.7 \\
EffiR-28MLP&2.8B&43.4 \\
\bottomrule
\end{tabular}
\caption{Comparison with the structural pruning method Sheared-LLaMA on LLaMA2-7B. Dropping 20 MLP layers is denoted as -20MLP; \textit{w/ $-30\%$} indicates reducing the width of the remaining MLP layers by 30\%. EffiR-20MLP \textit{w/ $-50\%$} achieves performance similar to Sheared-LLaMA at comparable model size, despite pruning only MLPs extremely aggressively (retaining \textasciitilde19\% of MLP parameters).}
\vspace{-3pt}
\label{tab:shearmed-llama}
\end{table}

We further compare \ours with another SOTA structural pruning method ShearedLlama~\cite{shearedllama}, which is also used for developing the Drama model~\cite{ma-etal-2025-drama}. For the sake of computation, we reuse the pruned LLaMA2 checkpoint released by the authors. We apply \ours directly to the LLaMA2 model.
Unlike \ours, which prunes only MLP layers, Sheared-LLaMA prunes all parameters, giving it greater pruning flexibility.

As shown in Table~\ref{tab:shearmed-llama}, the two-stage design of EffiR effectively mitigates the severe performance drop that occurs when applying layer dropping alone. At comparable scale, EffiR-20MLP \textit{w/ $-50\%$} (3.1B) achieves performance similar to Sheared-LLaMA (2.6B), despite restricting pruning to MLP layers only and using 10x less data for pruning-specific training. Notably, EffiR-20MLP \textit{w/ $-50\%$} retains just \textasciitilde19\% of the original MLP parameters, directly highlighting that a large fraction of MLP capacity is redundant for retrieval. 

This comparison also underscores the value of focusing on MLPs for retrieval. Sheared-LLaMA uses a pruning pipeline that compresses both attention and MLPs, yet the gap to MLP-only pruning with EffiR is small. We also emphasize that Sheared-LLaMA's pruning pipeline requires 0.4 billion training tokens from diverse domains. In contrast, EffiR achieves competitive performance using fewer tokens (\textasciitilde 0.03 billion). Thus, even if Sheared-LLaMA is slightly smaller in total parameters, EffiR offers a simpler alternative that isolates where the redundancy lies and recovers most of the benefit at a lower pruning cost.

\subsubsection{EffiR with Quantization}

We apply bitsandbytes NF4 quantization (with double quantization applied)~\cite{qlora} to both the full Mistral model and the EffiR-pruned models. The results in Table~\ref{tab:quantization} show that quantization behaves similarly for both the unmodified model and the EffiR-pruned model. In particular, training with EffiR does not degrade quantization performance, and in practice leads to substantial size reductions with minimal loss in effectiveness. This highlights the practical usefulness of EffiR as a technique that can be combined with quantization for even greater efficiency.

\begin{table}[h]
\small
\centering
\setlength{\tabcolsep}{3.5pt}
\setlength{\abovecaptionskip}{0.2cm}
\setlength{\belowcaptionskip}{-0.0cm}
\begin{tabular}{@{} l r r c @{}}
\toprule
Method & Precision & Size & BEIR Avg. \\
\midrule
\multirow{2}{*}{Full-Mistral-Model}
  & 16bit & 14.0 GB & 56.1 \\
  & 4bit  & 4.3 GB  & 56.0 \\
\midrule
\multirow{2}{*}{EffiR-Mistral-8MLP}
  & 16bit & 11.4 GB & 55.9 \\
  & 4bit  & 3.7 GB  & 55.7 \\
\midrule
\multirow{2}{*}{EffiR-Mistral-16MLP}
  & 16bit & 8.7 GB & 55.3 \\
  & 4bit  & 3.0 GB & 55.3 \\
\bottomrule
\end{tabular}
\caption{EffiR with quantization applied.}
\vspace{-10pt}
\label{tab:quantization}
\end{table}

\section{Conclusion}
In this work, we investigate the redundancy of LLMs as retrievers and examine the traditional pipeline for developing dense retrievers. Specifically, we reveal the parameter redundancy inherent in directly fine-tuning base models for retrieval tasks. To address this, we propose EffiR, a two-stage framework that compresses models in depth and width, followed by fine-tuning for retrieval tasks. \ours offers a general framework for developing retrieval models and provides valuable insights into efficient retriever design.

\section*{Limitations}
We acknowledge the following limitations:  
(i) English-centric evaluation: \ours is evaluated primarily on standard English retrieval benchmarks. Its effectiveness in multilingual or low-resource retrieval settings remains unexplored and may require further adaptation.  
(ii) Inference cost: Although \ours enables the development of efficient retrievers that preserve performance while reducing model size, the resulting EffiR models remain slower at inference compared to smaller architectures such as BERT-base.

\bibliography{custom}

@misc{gromov2024unreasonableineffectivenessdeeperlayers,
      title={The Unreasonable Ineffectiveness of the Deeper Layers}, 
      author={Andrey Gromov and Kushal Tirumala and Hassan Shapourian and Paolo Glorioso and Daniel A. Roberts},
      year={2024},
      eprint={2403.17887},
      archivePrefix={arXiv},
      primaryClass={cs.CL},
      url={https://arxiv.org/abs/2403.17887}, 
}

@misc{he2024matterstransformersattentionneeded,
      title={What Matters in Transformers? Not All Attention is Needed}, 
      author={Shwai He and Guoheng Sun and Zheyu Shen and Ang Li},
      year={2024},
      eprint={2406.15786},
      archivePrefix={arXiv},
      primaryClass={cs.LG},
      url={https://arxiv.org/abs/2406.15786}, 
}

@misc{geminiteam2024gemini,
      title={Gemini 1.5: Unlocking multimodal understanding across millions of tokens of context}, 
      author={Gemini},
      year={2024},
      eprint={2403.05530},
      archivePrefix={arXiv},
      primaryClass={cs.CL}
}

@misc{openai2024gpt4technicalreport,
      title={GPT-4 Technical Report}, 
      author={OpenAI and Josh Achiam and Steven Adler and Sandhini Agarwal and Lama Ahmad and Ilge Akkaya and Florencia Leoni Aleman and Diogo Almeida and Janko Altenschmidt and Sam Altman and Shyamal Anadkat and Red Avila and Igor Babuschkin and Suchir Balaji and Valerie Balcom and Paul Baltescu and Haiming Bao and Mohammad Bavarian and Jeff Belgum and Irwan Bello and Jake Berdine and Gabriel Bernadett-Shapiro and Christopher Berner and Lenny Bogdonoff and Oleg Boiko and Madelaine Boyd and Anna-Luisa Brakman and Greg Brockman and Tim Brooks and Miles Brundage and Kevin Button and Trevor Cai and Rosie Campbell and Andrew Cann and Brittany Carey and Chelsea Carlson and Rory Carmichael and Brooke Chan and Che Chang and Fotis Chantzis and Derek Chen and Sully Chen and Ruby Chen and Jason Chen and Mark Chen and Ben Chess and Chester Cho and Casey Chu and Hyung Won Chung and Dave Cummings and Jeremiah Currier and Yunxing Dai and Cory Decareaux and Thomas Degry and Noah Deutsch and Damien Deville and Arka Dhar and David Dohan and Steve Dowling and Sheila Dunning and Adrien Ecoffet and Atty Eleti and Tyna Eloundou and David Farhi and Liam Fedus and Niko Felix and Simón Posada Fishman and Juston Forte and Isabella Fulford and Leo Gao and Elie Georges and Christian Gibson and Vik Goel and Tarun Gogineni and Gabriel Goh and Rapha Gontijo-Lopes and Jonathan Gordon and Morgan Grafstein and Scott Gray and Ryan Greene and Joshua Gross and Shixiang Shane Gu and Yufei Guo and Chris Hallacy and Jesse Han and Jeff Harris and Yuchen He and Mike Heaton and Johannes Heidecke and Chris Hesse and Alan Hickey and Wade Hickey and Peter Hoeschele and Brandon Houghton and Kenny Hsu and Shengli Hu and Xin Hu and Joost Huizinga and Shantanu Jain and Shawn Jain and Joanne Jang and Angela Jiang and Roger Jiang and Haozhun Jin and Denny Jin and Shino Jomoto and Billie Jonn and Heewoo Jun and Tomer Kaftan and Łukasz Kaiser and Ali Kamali and Ingmar Kanitscheider and Nitish Shirish Keskar and Tabarak Khan and Logan Kilpatrick and Jong Wook Kim and Christina Kim and Yongjik Kim and Jan Hendrik Kirchner and Jamie Kiros and Matt Knight and Daniel Kokotajlo and Łukasz Kondraciuk and Andrew Kondrich and Aris Konstantinidis and Kyle Kosic and Gretchen Krueger and Vishal Kuo and Michael Lampe and Ikai Lan and Teddy Lee and Jan Leike and Jade Leung and Daniel Levy and Chak Ming Li and Rachel Lim and Molly Lin and Stephanie Lin and Mateusz Litwin and Theresa Lopez and Ryan Lowe and Patricia Lue and Anna Makanju and Kim Malfacini and Sam Manning and Todor Markov and Yaniv Markovski and Bianca Martin and Katie Mayer and Andrew Mayne and Bob McGrew and Scott Mayer McKinney and Christine McLeavey and Paul McMillan and Jake McNeil and David Medina and Aalok Mehta and Jacob Menick and Luke Metz and Andrey Mishchenko and Pamela Mishkin and Vinnie Monaco and Evan Morikawa and Daniel Mossing and Tong Mu and Mira Murati and Oleg Murk and David Mély and Ashvin Nair and Reiichiro Nakano and Rajeev Nayak and Arvind Neelakantan and Richard Ngo and Hyeonwoo Noh and Long Ouyang and Cullen O'Keefe and Jakub Pachocki and Alex Paino and Joe Palermo and Ashley Pantuliano and Giambattista Parascandolo and Joel Parish and Emy Parparita and Alex Passos and Mikhail Pavlov and Andrew Peng and Adam Perelman and Filipe de Avila Belbute Peres and Michael Petrov and Henrique Ponde de Oliveira Pinto and Michael and Pokorny and Michelle Pokrass and Vitchyr H. Pong and Tolly Powell and Alethea Power and Boris Power and Elizabeth Proehl and Raul Puri and Alec Radford and Jack Rae and Aditya Ramesh and Cameron Raymond and Francis Real and Kendra Rimbach and Carl Ross and Bob Rotsted and Henri Roussez and Nick Ryder and Mario Saltarelli and Ted Sanders and Shibani Santurkar and Girish Sastry and Heather Schmidt and David Schnurr and John Schulman and Daniel Selsam and Kyla Sheppard and Toki Sherbakov and Jessica Shieh and Sarah Shoker and Pranav Shyam and Szymon Sidor and Eric Sigler and Maddie Simens and Jordan Sitkin and Katarina Slama and Ian Sohl and Benjamin Sokolowsky and Yang Song and Natalie Staudacher and Felipe Petroski Such and Natalie Summers and Ilya Sutskever and Jie Tang and Nikolas Tezak and Madeleine B. Thompson and Phil Tillet and Amin Tootoonchian and Elizabeth Tseng and Preston Tuggle and Nick Turley and Jerry Tworek and Juan Felipe Cerón Uribe and Andrea Vallone and Arun Vijayvergiya and Chelsea Voss and Carroll Wainwright and Justin Jay Wang and Alvin Wang and Ben Wang and Jonathan Ward and Jason Wei and CJ Weinmann and Akila Welihinda and Peter Welinder and Jiayi Weng and Lilian Weng and Matt Wiethoff and Dave Willner and Clemens Winter and Samuel Wolrich and Hannah Wong and Lauren Workman and Sherwin Wu and Jeff Wu and Michael Wu and Kai Xiao and Tao Xu and Sarah Yoo and Kevin Yu and Qiming Yuan and Wojciech Zaremba and Rowan Zellers and Chong Zhang and Marvin Zhang and Shengjia Zhao and Tianhao Zheng and Juntang Zhuang and William Zhuk and Barret Zoph},
      year={2024},
      eprint={2303.08774},
      archivePrefix={arXiv},
      primaryClass={cs.CL},
      url={https://arxiv.org/abs/2303.08774}, 
}

@misc{jiang2023mistral7b,
      title={Mistral 7B}, 
      author={Albert Q. Jiang and Alexandre Sablayrolles and Arthur Mensch and Chris Bamford and Devendra Singh Chaplot and Diego de las Casas and Florian Bressand and Gianna Lengyel and Guillaume Lample and Lucile Saulnier and Lélio Renard Lavaud and Marie-Anne Lachaux and Pierre Stock and Teven Le Scao and Thibaut Lavril and Thomas Wang and Timothée Lacroix and William El Sayed},
      year={2023},
      eprint={2310.06825},
      archivePrefix={arXiv},
      primaryClass={cs.CL},
      url={https://arxiv.org/abs/2310.06825}, 
}

@inproceedings{chen-etal-2024-m3,
    title = "{M}3-Embedding: Multi-Linguality, Multi-Functionality, Multi-Granularity Text Embeddings Through Self-Knowledge Distillation",
    author = "Chen, Jianlyu  and
      Xiao, Shitao  and
      Zhang, Peitian  and
      Luo, Kun  and
      Lian, Defu  and
      Liu, Zheng",
    editor = "Ku, Lun-Wei  and
      Martins, Andre  and
      Srikumar, Vivek",
    booktitle = "Findings of the Association for Computational Linguistics: ACL 2024",
    month = aug,
    year = "2024",
    address = "Bangkok, Thailand",
    publisher = "Association for Computational Linguistics",
    url = "https://aclanthology.org/2024.findings-acl.137/",
    doi = "10.18653/v1/2024.findings-acl.137",
    pages = "2318--2335"
}

@article{Wang2022TextEB,
  title={Text Embeddings by Weakly-Supervised Contrastive Pre-training},
  author={Liang Wang and Nan Yang and Xiaolong Huang and Binxing Jiao and Linjun Yang and Daxin Jiang and Rangan Majumder and Furu Wei},
  journal={ArXiv},
  year={2022},
  volume={abs/2212.03533},
  url={https://api.semanticscholar.org/CorpusID:254366618}
}

@misc{li2024makingtextembeddersfewshot,
      title={Making Text Embedders Few-Shot Learners}, 
      author={Chaofan Li and MingHao Qin and Shitao Xiao and Jianlyu Chen and Kun Luo and Yingxia Shao and Defu Lian and Zheng Liu},
      year={2024},
      eprint={2409.15700},
      archivePrefix={arXiv},
      primaryClass={cs.IR},
      url={https://arxiv.org/abs/2409.15700}, 
}

@inproceedings{karpukhin2020dense,
  title={Dense Passage Retrieval for Open-Domain Question Answering},
  author={Karpukhin, Vladimir and Oguz, Barlas and Min, Sewon and Lewis, Patrick and Wu, Ledell and Edunov, Sergey and Chen, Danqi and Yih, Wen-tau},
  booktitle={EMNLP},
  year={2020}
}

@inproceedings{xiong2021approximate,
  title={Approximate nearest neighbor negative contrastive learning for dense text retrieval},
  author={Xiong, Lee and Wu, Chenyan and Xiong, Ye and Luan, Jian and Rogriguez, Keith and Zettlemoyer, Luke and Sun, Mingyang},
  booktitle={ICLR},
  year={2021}
}

@inproceedings{thakur2021beir,
  title={BEIR: A Heterogeneous Benchmark for Zero-shot Evaluation of Information Retrieval Models},
  author={Thakur, Nandan and Reimers, Nils and Sanii, Andreas and Gurevych, Iryna},
  booktitle={NeurIPS},
  year={2021}
}

@misc{muennighoff2023mtebmassivetextembedding,
      title={MTEB: Massive Text Embedding Benchmark}, 
      author={Niklas Muennighoff and Nouamane Tazi and Loïc Magne and Nils Reimers},
      year={2023},
      eprint={2210.07316},
      archivePrefix={arXiv},
      primaryClass={cs.CL},
      url={https://arxiv.org/abs/2210.07316}, 
}

@inproceedings{wang2024improving-text,
    title = "Improving Text Embeddings with Large Language Models",
    author = "Wang, Liang  and
      Yang, Nan  and
      Huang, Xiaolong  and
      Yang, Linjun  and
      Majumder, Rangan  and
      Wei, Furu",
    booktitle = "Proceedings of the 62nd Annual Meeting of the Association for Computational Linguistics (Volume 1: Long Papers)",
    month = aug,
    year = "2024",
    address = "Bangkok, Thailand",
    publisher = "Association for Computational Linguistics",
    url = "https://aclanthology.org/2024.acl-long.642/",
    pages = "11897--11916",
}

@inproceedings{
siddiqui2024adeeperlook,
title={A deeper look at depth pruning of {LLM}s},
author={Shoaib Ahmed Siddiqui and Xin Dong and Greg Heinrich and Thomas Breuel and Jan Kautz and David Krueger and Pavlo Molchanov},
booktitle={ICML 2024 Workshop on Theoretical Foundations of Foundation Models},
year={2024},
url={https://openreview.net/forum?id=9B7ayWclwN}
}

@article{raffel2020exploring,
author = {Raffel, Colin and Shazeer, Noam and Roberts, Adam and Lee, Katherine and Narang, Sharan and Matena, Michael and Zhou, Yanqi and Li, Wei and Liu, Peter J.},
title = {Exploring the limits of transfer learning with a unified text-to-text transformer},
year = {2020},
publisher = {JMLR.org},
volume = {21},
number = {1},
journal = {J. Mach. Learn. Res.},
month = jan,
articleno = {140},
numpages = {67},
}

@inproceedings{ma2024finetuningllama,
author = {Ma, Xueguang and Wang, Liang and Yang, Nan and Wei, Furu and Lin, Jimmy},
title = {Fine-Tuning LLaMA for Multi-Stage Text Retrieval},
year = {2024},
publisher = {Association for Computing Machinery},
address = {New York, NY, USA},
url = {https://doi.org/10.1145/3626772.3657951},
booktitle = {Proceedings of the 47th International ACM SIGIR Conference on Research and Development in Information Retrieval},
pages = {2421–2425},
location = {Washington DC, USA},
series = {SIGIR '24}
}

@article{msmarco,
      title={MS MARCO: A Human Generated MAchine Reading COmprehension Dataset}, 
      author={Payal Bajaj and Daniel Campos and Nick Craswell and Li Deng and Jianfeng Gao and Xiaodong Liu and Rangan Majumder and Andrew McNamara and Bhaskar Mitra and Tri Nguyen and Mir Rosenberg and Xia Song and Alina Stoica and Saurabh Tiwary and Tong Wang},
      journal={arXiv preprint arXiv:1611.09268},
      year={2018},
      url={https://arxiv.org/abs/1611.09268}, 
}

@inproceedings{devlin-etal-2019-bert,
    title = "{BERT}: Pre-training of Deep Bidirectional Transformers for Language Understanding",
    author = "Devlin, Jacob  and
      Chang, Ming-Wei  and
      Lee, Kenton  and
      Toutanova, Kristina",
    editor = "Burstein, Jill  and
      Doran, Christy  and
      Solorio, Thamar",
    booktitle = "Proceedings of the 2019 Conference of the North {A}merican Chapter of the Association for Computational Linguistics: Human Language Technologies, Volume 1 (Long and Short Papers)",
    month = jun,
    year = "2019",
    address = "Minneapolis, Minnesota",
    publisher = "Association for Computational Linguistics",
    url = "https://aclanthology.org/N19-1423/",
    doi = "10.18653/v1/N19-1423",
    pages = "4171--4186"
}

@article{
izacard2022unsupervised,
title={Unsupervised Dense Information Retrieval with Contrastive Learning},
author={Gautier Izacard and Mathilde Caron and Lucas Hosseini and Sebastian Riedel and Piotr Bojanowski and Armand Joulin and Edouard Grave},
journal={Transactions on Machine Learning Research},
issn={2835-8856},
year={2022},
url={https://openreview.net/forum?id=jKN1pXi7b0},
note={}
}

@inproceedings{hofst2021efficiently,
author = {Hofst\"{a}tter, Sebastian and Lin, Sheng-Chieh and Yang, Jheng-Hong and Lin, Jimmy and Hanbury, Allan},
title = {Efficiently Teaching an Effective Dense Retriever with Balanced Topic Aware Sampling},
year = {2021},
isbn = {9781450380379},
publisher = {Association for Computing Machinery},
address = {New York, NY, USA},
url = {https://doi.org/10.1145/3404835.3462891},
doi = {10.1145/3404835.3462891},
booktitle = {Proceedings of the 44th International ACM SIGIR Conference on Research and Development in Information Retrieval},
pages = {113–122},
numpages = {10},
location = {Virtual Event, Canada},
series = {SIGIR '21}
}

@inproceedings{
weller2025promptriever,
title={Promptriever: Instruction-Trained Retrievers Can Be Prompted Like Language Models},
author={Orion Weller and Benjamin Van Durme and Dawn Lawrie and Ashwin Paranjape and Yuhao Zhang and Jack Hessel},
booktitle={The Thirteenth International Conference on Learning Representations},
year={2025},
url={https://openreview.net/forum?id=odvSjn416y}
}

@misc{kusupati2024matryoshkarepresentationlearning,
      title={Matryoshka Representation Learning}, 
      author={Aditya Kusupati and Gantavya Bhatt and Aniket Rege and Matthew Wallingford and Aditya Sinha and Vivek Ramanujan and William Howard-Snyder and Kaifeng Chen and Sham Kakade and Prateek Jain and Ali Farhadi},
      year={2024},
      eprint={2205.13147},
      archivePrefix={arXiv},
      primaryClass={cs.LG},
      url={https://arxiv.org/abs/2205.13147}, 
}

@article{
kumar2024ehi,
title={{EHI}: End-to-end Learning of Hierarchical Index for Efficient Dense Retrieval},
author={Ramnath Kumar and Anshul Mittal and Nilesh Gupta and Aditya Kusupati and Inderjit S Dhillon and Prateek Jain},
journal={Transactions on Machine Learning Research},
issn={2835-8856},
year={2024},
url={https://openreview.net/forum?id=GeLLOGsHV9},
note={}
}

@inproceedings{luo-etal-2024-large,
    title = "Large Language Models as Foundations for Next-Gen Dense Retrieval: A Comprehensive Empirical Assessment",
    author = "Luo, Kun  and
      Qin, Minghao  and
      Liu, Zheng  and
      Xiao, Shitao  and
      Zhao, Jun  and
      Liu, Kang",
    editor = "Al-Onaizan, Yaser  and
      Bansal, Mohit  and
      Chen, Yun-Nung",
    booktitle = "Proceedings of the 2024 Conference on Empirical Methods in Natural Language Processing",
    month = nov,
    year = "2024",
    address = "Miami, Florida, USA",
    publisher = "Association for Computational Linguistics",
    url = "https://aclanthology.org/2024.emnlp-main.80/",
    doi = "10.18653/v1/2024.emnlp-main.80",
    pages = "1354--1365"
}

@inproceedings{sun-etal-2024-mair,
    title = "{MAIR}: A Massive Benchmark for Evaluating Instructed Retrieval",
    author = "Sun, Weiwei  and
      Shi, Zhengliang  and
      Long, Wu Jiu  and
      Yan, Lingyong  and
      Ma, Xinyu  and
      Liu, Yiding  and
      Cao, Min  and
      Yin, Dawei  and
      Ren, Zhaochun",
    editor = "Al-Onaizan, Yaser  and
      Bansal, Mohit  and
      Chen, Yun-Nung",
    booktitle = "Proceedings of the 2024 Conference on Empirical Methods in Natural Language Processing",
    month = nov,
    year = "2024",
    address = "Miami, Florida, USA",
    publisher = "Association for Computational Linguistics",
    url = "https://aclanthology.org/2024.emnlp-main.778/",
    doi = "10.18653/v1/2024.emnlp-main.778",
    pages = "14044--14067"
}

@inproceedings{meng2022locating,
author = {Meng, Kevin and Bau, David and Andonian, Alex and Belinkov, Yonatan},
title = {Locating and editing factual associations in GPT},
year = {2022},
isbn = {9781713871088},
publisher = {Curran Associates Inc.},
address = {Red Hook, NY, USA},
booktitle = {Proceedings of the 36th International Conference on Neural Information Processing Systems},
articleno = {1262},
numpages = {14},
location = {New Orleans, LA, USA},
series = {NIPS '22}
}

@misc{zhu2020modifyingmemoriestransformermodels,
      title={Modifying Memories in Transformer Models}, 
      author={Chen Zhu and Ankit Singh Rawat and Manzil Zaheer and Srinadh Bhojanapalli and Daliang Li and Felix Yu and Sanjiv Kumar},
      year={2020},
      eprint={2012.00363},
      archivePrefix={arXiv},
      primaryClass={cs.CL},
      url={https://arxiv.org/abs/2012.00363}, 
}

@inproceedings{lei-etal-2023-unsupervised,
    title = "Unsupervised Dense Retrieval with Relevance-Aware Contrastive Pre-Training",
    author = "Lei, Yibin  and
      Ding, Liang  and
      Cao, Yu  and
      Zan, Changtong  and
      Yates, Andrew  and
      Tao, Dacheng",
    booktitle = "Findings of the Association for Computational Linguistics: ACL 2023",
    year = "2023",
    publisher = "Association for Computational Linguistics",
    url = "https://aclanthology.org/2023.findings-acl.695/",
}

@article{robertson1995okapi,
  title={Okapi at TREC-3},
  author={Robertson, Stephen E and Walker, Steve and Jones, Susan and Hancock-Beaulieu, Micheline M and Gatford, Mike and others},
  journal={Nist Special Publication Sp},
  volume={109},
  pages={109},
  year={1995},
  publisher={National Instiute of Standards \& Technology}
}

@inproceedings{lei-etal-2025-enhancing,
    title = "Enhancing Lexicon-Based Text Embeddings with Large Language Models",
    author = "Lei, Yibin  and
      Shen, Tao  and
      Cao, Yu  and
      Yates, Andrew",
    booktitle = "Proceedings of the 63rd Annual Meeting of the Association for Computational Linguistics (Volume 1: Long Papers)",
    month = jul,
    year = "2025",
    address = "Vienna, Austria",
    publisher = "Association for Computational Linguistics",
    url = "https://aclanthology.org/2025.acl-long.930/",
    pages = "18986--19001",
}

@inproceedings{sesmic,
author = {Bruch, Sebastian and Nardini, Franco Maria and Rulli, Cosimo and Venturini, Rossano},
title = {Efficient Inverted Indexes for Approximate Retrieval over Learned Sparse Representations},
year = {2024},
publisher = {Association for Computing Machinery},
address = {New York, NY, USA},
url = {https://doi.org/10.1145/3626772.3657769},
booktitle = {Proceedings of the 47th International ACM SIGIR Conference on Research and Development in Information Retrieval},
pages = {152–162},
location = {Washington DC, USA},
series = {SIGIR '24}
}

@inproceedings{ma-etal-2025-drama,
    title = "{DRAMA}: Diverse Augmentation from Large Language Models to Smaller Dense Retrievers",
    author = "Ma, Xueguang  and
      Lin, Xi Victoria  and
      Oguz, Barlas  and
      Lin, Jimmy  and
      Yih, Wen-tau  and
      Chen, Xilun",
    booktitle = "Proceedings of the 63rd Annual Meeting of the Association for Computational Linguistics (Volume 1: Long Papers)",
    month = jul,
    year = "2025",
    address = "Vienna, Austria",
    publisher = "Association for Computational Linguistics",
    url = "https://aclanthology.org/2025.acl-long.1457/",
    pages = "30170--30186",
}

@inproceedings{
sun2024awanda,
title={A Simple and Effective Pruning Approach for Large Language Models},
author={Mingjie Sun and Zhuang Liu and Anna Bair and J Zico Kolter},
booktitle={The Twelfth International Conference on Learning Representations},
year={2024},
url={https://openreview.net/forum?id=PxoFut3dWW}
}

@article{he2025understandingharnessingsparsityunified,
      title={Understanding and Harnessing Sparsity in Unified Multimodal Models}, 
      author={Shwai He and Chaorui Deng and Ang Li and Shen Yan},
      journal={arXiv preprint arXiv:2512.02351},
      year={2025},
      url={https://arxiv.org/abs/2512.02351}, 
}

@article{starbucks,
      title={Starbucks-v2: Improved Training for 2D Matryoshka Embeddings}, 
      author={Shengyao Zhuang and Shuai Wang and Fabio Zheng and Bevan Koopman and Guido Zuccon},
      journal={arXiv preprint arXiv:2410.13230},
      year={2025},
      url={https://arxiv.org/abs/2410.13230}, 
}

@inproceedings{
espresso,
title={{ESE}: Espresso Sentence Embeddings},
author={Xianming LI and Zongxi Li and Jing Li and Haoran Xie and Qing Li},
booktitle={The Thirteenth International Conference on Learning Representations},
year={2025},
url={https://openreview.net/forum?id=plgLA2YBLH}
}

@inproceedings{sparsegpt,
author = {Frantar, Elias and Alistarh, Dan},
title = {SparseGPT: massive language models can be accurately pruned in one-shot},
year = {2023},
publisher = {JMLR.org},
series = {ICML'23}
}

@inproceedings{
shearedllama,
title={Sheared {LL}a{MA}: Accelerating Language Model Pre-training via Structured Pruning},
author={Mengzhou Xia and Tianyu Gao and Zhiyuan Zeng and Danqi Chen},
booktitle={The Twelfth International Conference on Learning Representations},
year={2024},
url={https://openreview.net/forum?id=09iOdaeOzp}
}

@inproceedings{
qlora,
title={{QL}o{RA}: Efficient Finetuning of Quantized {LLM}s},
author={Tim Dettmers and Artidoro Pagnoni and Ari Holtzman and Luke Zettlemoyer},
booktitle={Thirty-seventh Conference on Neural Information Processing Systems},
year={2023},
}

@article{grattafiori2024llama3herdmodels,
      title={The Llama 3 Herd of Models}, 
      author={Aaron Grattafiori and Abhimanyu Dubey and Abhinav Jauhri and Abhinav Pandey and Abhishek Kadian and others},
      journal={arXiv preprint arXiv:2407.21783},
      year={2024},
      url={https://arxiv.org/abs/2407.21783}, 
}

@article{gemmateam2024gemma2improvingopen,
      title={Gemma 2: Improving Open Language Models at a Practical Size}, 
      author={Gemma Team and Morgane Riviere and Shreya Pathak and Pier Giuseppe Sessa and Cassidy Hardin and others},
      journal={arXiv preprint arXiv:2408.00118},
      year={2024},
      url={https://arxiv.org/abs/2408.00118}, 
}

@article{bai2023qwen1.5technicalreport,
      title={Qwen Technical Report}, 
      author={Jinze Bai and Shuai Bai and Yunfei Chu and Zeyu Cui and Kai Dang and others},
      journal={arXiv preprint arXiv:2309.16609},
      year={2023},
      url={https://arxiv.org/abs/2309.16609}, 
}

\clearpage
\newpage
\appendix
\section{Appendix}

\subsection{Sigmoid Surrogate for $\ell_0$-norm}
\label{app:relaxation}
To encourage sparsity during training without enforcing hard thresholding, we adopt a simple and efficient sigmoid-based surrogate inspired by the $\ell_0$-norm. Specifically, for a set of parameters $\{x_i\}$, the $\ell_0$-norm counts the number of non-zero entries:
\[
\|\mathbf{x}\|_0 = \sum_{i=1}^d \mathbb{I}[x_i \ne 0].
\]
This objective is discontinuous and therefore incompatible with standard gradient-based optimization. We therefore use a differentiable surrogate based on the sigmoid of the parameter magnitude:
\[
\tilde{\mathcal{R}}(\mathbf{x}) = \sum_{i=1}^d \sigma(\beta |x_i|),
\]
where $\sigma(\cdot)$ denotes the sigmoid function and $\beta>0$ controls the sharpness of the transition. 
Note that $\sigma(0)=0.5$, so this surrogate is not a direct approximation of $\mathbb{I}[x_i \neq 0]$; rather, when minimized, it softly encourages parameters to shrink toward zero by assigning the lowest penalty to zero-valued entries and increasingly larger penalties to non-zero magnitudes.
 Increasing $\beta$ sharpens the transition around the origin (i.e., it more strongly separates small from large magnitudes), but overly large values may harm gradient stability; in practice, we fix $\beta = 5.0$.

\subsection{Training Configurations}
\label{app:training_configutations}
We train each model for one epoch using LoRA, applied to the \texttt{v\_proj}, \texttt{q\_proj}, \texttt{k\_proj}, \texttt{gate\_proj}, \texttt{down\_proj}, \texttt{o\_proj}, \texttt{up\_proj} layers, with a rank of 32 and $\alpha = 64$. The learning rate is set to 1e-4, and each sample consists of one positive and seven negatives. We also include in-batch negatives and apply the KL-divergence loss to distill ranking scores from the BGE-reranker. For the self-slimming setup, we fine-tune the scaling factors for 500 steps using full-parameter training and apply a regularization weight $\lambda$ of 1e-8. Afterward, we prune 30\% of the intermediate dimensions in the MLP layers based on their learned scaling factors, removing the dimensions with the lowest values across all layers. 

\newpage
\subsection{BEIR Statisitcs}
We provide the detailed BEIR statistics in Table~\ref{tab:data_stat_retrievaaaal}.
\begin{table}[hbt]
\setlength{\abovecaptionskip}{0.1cm}
\setlength{\belowcaptionskip}{-0.2cm}
\centering
\small
\setlength{\tabcolsep}{3pt}
\begin{tabular}{@{} l rr @{}}
\toprule
Dataset & ~~~~~~~\#Test & ~~~~~~~\#Corpus \\
\midrule
Scifact & 300 & 5,183 \\
Arguana & 1,406 & 8,674 \\
Trec-Covid  & 50 & 171,332 \\
FiQA-2018   & 648 & 57,638 \\
DBPedia    & 400 & 4,635,922 \\
NFCorpus & 323 & 3,633 \\
NQ & 3,452 & 2,681,468\\
HotpotQA & 7,405 & 5,233,329 \\
Touche-2020 & 49 &  382,545 \\
Quora & 10,000 &  522,931 \\
SCIDOCS & 1,000 &  25,657 \\
FEVER & 6,666 & 5,416,568 \\
Climate-FEVER & 1,535 &  5,416,593 \\
\bottomrule
\end{tabular}
\caption{Dataset Statistics}
\label{tab:data_stat_retrievaaaal}
\end{table}

\subsection{Additional Layer Dropping Results}
\label{app:llama3}

We present the results of coarse-grained layer dropping analysis on the LLaMA3-8B, Qwen-2.5-1.5B, Qwen-2.5-3B, Qwen-2.5-7B, and ModernBERT-base in Tables~\ref{table:llama3-8b}, \ref{table:qwen2.5-1.5B}, \ref{table:qwen2.5-3B}, \ref{table:qwen2.5-7B}, and \ref{table:modernbert-base}, with contrastive learning applied. Similar to Mistral-7B, dropping MLP layers results in a smaller performance degradation compared to pruning attention layers or entire transformer blocks across these models.

\subsection{Results on Retrieval Latency}
We find that different EffiR variants and the original model, which have the same embedding dimension, achieve comparable retrieval latency: \textasciitilde2.28 ms/query on the NFCorpus dataset using the Faiss Flat index on a single core of the AMD EPYC 7763 CPU. This indicates that EffiR does not introduce any retrieval-side overhead.

\subsection{Results on Embedding Space Alignment}
 We compute the embedding-space isotropy values (based on average pairwise cosine similarity) for both the original full model and EffiR-20MLP \textit{w/ $-20\%$} using 10,000 MS MARCO queries. The two models obtain comparable isotropy scores (0.28 vs. 0.34), suggesting that EffiR largely preserves the geometric structure of the embedding space.

\subsection{Dropping-Order Heat Maps}
\begin{figure*}[h]
  \centering
  \begin{subfigure}[t]{0.31\linewidth}
    \centering
    \includegraphics[width=\linewidth]{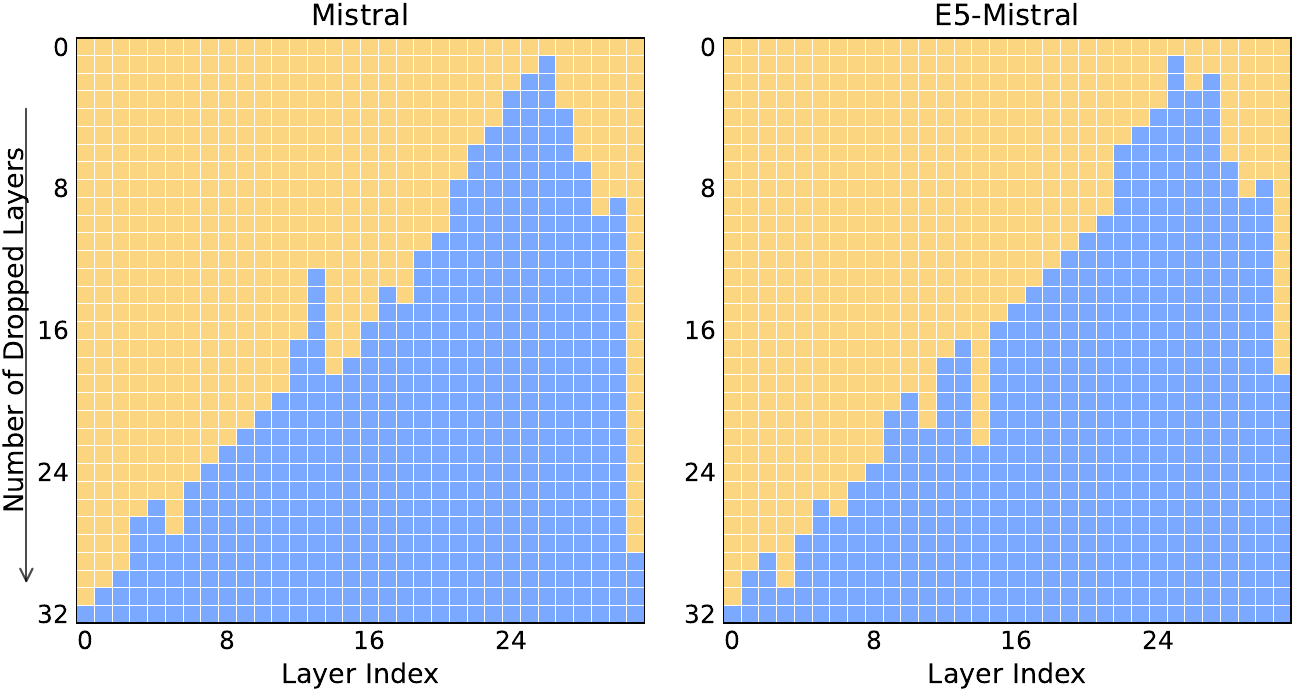}
    \caption{Block}
    \label{fig:drop_order_block_full}
  \end{subfigure}
  \hfill
  \begin{subfigure}[t]{0.31\linewidth}
    \centering
    \includegraphics[width=\linewidth]{figs/drop_order_MLP.pdf}
    \caption{MLP}
    \label{fig:drop_order_mlp_full}
  \end{subfigure}
  \hfill
  \begin{subfigure}[t]{0.31\linewidth}
    \centering
    \includegraphics[width=\linewidth]{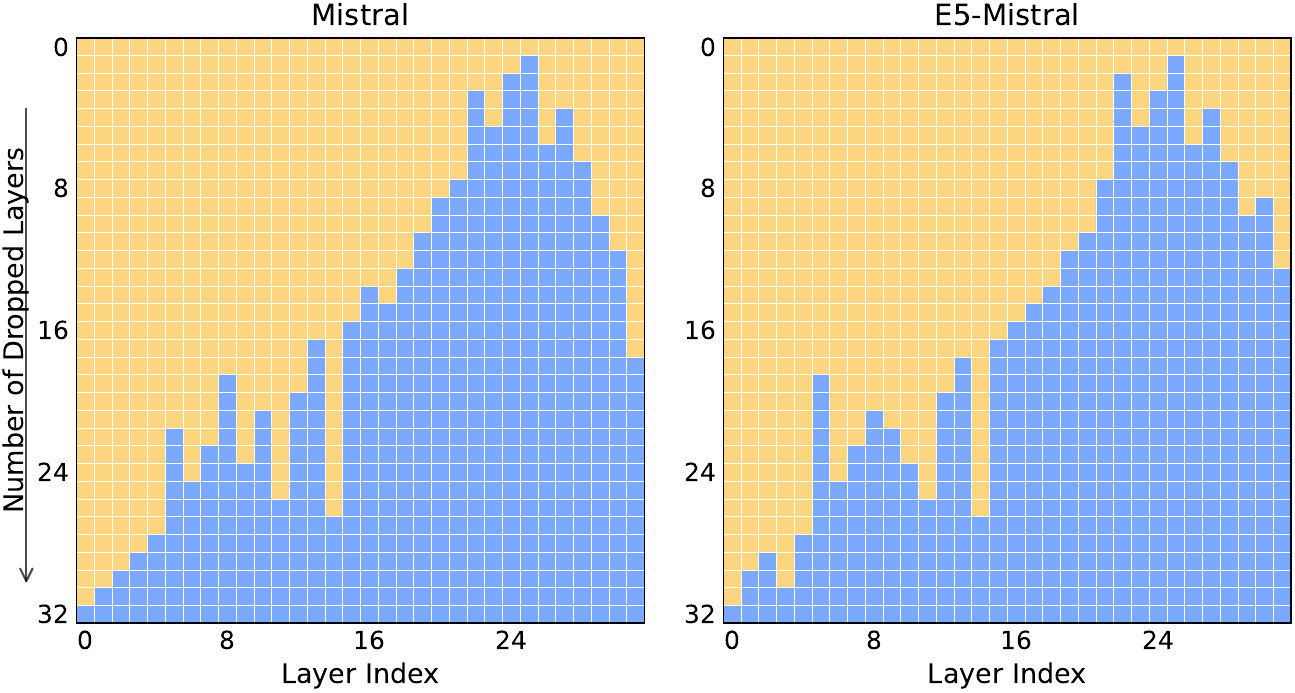}
    \caption{Attention}
    \label{fig:drop_order_attn_full}
  \end{subfigure}

  \caption{\textbf{Dropping‑order heat maps} for the base model (Mistral‑7B) and its embedding variant (E5‑Mistral).}
  \label{fig:drop_order_full}
\end{figure*}

Figure~\ref{fig:drop_order_full} presents drop-order heat maps visualizing the layer-wise redundancy of attention layers and Transformer blocks in both Mistral-7B and its retrieval-tuned variant, E5-Mistral. As with MLP layers, the later layers tend to be more redundant and thus more prunable.

\subsection{Why Attention is Critical: a Case Study}

We have shown that attention layers are more critical in embedding models than in generative models. To further explore this, we examine how selectively dropping MLP or attention layers impacts retrieval performance.

\begin{table*}[h]
\centering
\scalebox{0.55}{
\begin{tabular}{l|r|rr|rr|rrrr|r}
\toprule
 \textbf{\textit{LLAMA3-8B}} & Full-Model & \ours-8A & \ours-16A & \ours-8B & \ours-16B & \ours-8M & \ours-16M & \ours-20M & \ours-24M & \ours-16M8A \\
 \midrule
Arguana       & 48.6  & 46.5 & 40.2 & 50.4 & 37.5 & 51.5 & 52.9 & 42.1 & 42.1 & 44.5 \\
Climate-FEVER & 31.0 & 29.1 & 26.5 & 30.5 & 27.3 & 29.4 & 28.9 & 29.6 & 21.1 & 28.8 \\
DBPedia       & 43.7  & 43.0 & 35.0 & 38.9 & 34.2 & 43.9 & 43.1 & 43.8 & 33.6 & 39.1 \\
FEVER         & 83.4  & 81.7 & 79.1 & 80.9 & 78.4 & 83.2 & 83.3 & 84.5 & 68.1 & 81.2 \\
FiQA          & 45.8 & 41.9 & 34.9 & 40.6 & 32.1 & 42.4 & 41.4 & 39.1 & 31.5 & 39.1 \\
HotpotQA      & 68.5  & 67.5 & 58.5 & 66.9 & 61.2 & 70.5 & 68.6 & 67.3 & 57.5 & 65.3 \\
NFCorpus      & 37.8  & 33.8 & 32.0 & 31.1 & 28.1 & 35.4 & 34.9 & 35.5 & 30.9 & 35.1 \\
NQ            & 62.4  & 63.9 & 58.1 & 62.1 & 56.5 & 63.8 & 62.4 & 61.7 & 49.4 & 61.0 \\
Quora         & 86.8  & 86.5 & 85.9 & 87.9 & 85.9 & 87.0 & 87.1 & 86.7 & 83.7 & 86.8 \\
SCIDOCS       & 18.1  & 15.8 & 12.7 & 14.8 & 12.3 & 17.6 & 15.8 & 16.4 & 15.3 & 15.8 \\
SciFact       & 75.6  & 73.3 & 69.0 & 73.2 & 69.0 & 74.5 & 72.8 & 73.5 & 64.7 & 73.8 \\
TREC-COVID    & 84.7  & 81.1 & 78.9 & 79.7 & 76.3 & 82.1 & 83.1 & 81.2 & 75.5 & 82.9 \\
Touche-2020   & 30.5  & 30.3 & 25.4 & 29.5 & 24.4 &  32.6   & 31.4 & 31.6 & 26.3 & 26.6 \\
\midrule
Average       &  55.5 & 53.4 & 48.9 & 52.8 & 47.9 & 54.9 & 54.3 & 53.3 & 46.1 & 52.3  \\
\bottomrule
\end{tabular}
}
\vspace{0.2cm}
\caption{Effectiveness (nDCG@10) of LLAMA3-8B variants trained after coarse-grained layer dropping across different architectural components. We compare pruning of MLP layers (e.g., \ours-16M), attention layers (e.g., \ours-16A), full transformer blocks (e.g., \ours-16B), and their combinations (e.g., \ours-16M8A).  
} 
\label{table:llama3-8b}
\end{table*}

\begin{table*}[h]
\centering
\scalebox{0.55}{
\begin{tabular}{l|r|rr|rrr}
\toprule
 \textbf{\textit{Qwen2.5-1.5B}} & Full-Model & \ours-8A & \ours-16A & \ours-8M & \ours-16M & \ours-20M \\
 \midrule
Arguana       &  53.8 & 50.5 & 41.5 & 52.0 & 49.1 & 43.9\\
Climate-FEVER &  26.6 & 25.5 & 20.3 & 25.7 & 25.2 & 21.2\\
DBPedia       &  42.0 & 38.4 & 31.3 & 40.6 & 38.2 & 32.4\\ 
FEVER         &  78.3 & 73.0 & 66.2 & 79.6 & 77.9 & 73.7\\ 
FiQA          &  38.5 & 35.0 & 26.2 & 37.3 & 34.0 & 30.2\\
HotpotQA      &  64.1 & 59.8 & 41.8 & 63.3 & 60.5 & 55.8\\
NFCorpus      &  35.7 & 36.1 & 32.1 & 35.6 & 35.8 & 30.1\\ 
NQ            &  57.8 & 55.1 & 44.8 & 55.8 & 53.6 & 46.2\\ 
Quora         &  86.4 & 86.9 & 83.6 & 86.8 & 84.8 & 82.6\\
SCIDOCS       &  17.8 & 16.9 & 12.9 & 17.5 & 15.4 & 13.3\\
SciFact       &  73.0 & 69.3 & 56.5 & 70.6 & 69.3 & 64.8\\ 
TREC-COVID    &  82.8 & 82.7 & 68.7 & 83.2 & 83.6 & 77.9\\ 
Touche-2020   &  30.7 & 26.3 & 24.6 & 28.9 & 29.8 & 29.9\\ 
\midrule  
Average       &  52.9 & 50.4 & 42.3 & 52.1 & 50.6 & 46.3\\ 
\bottomrule
\end{tabular}
}
\vspace{0.2cm}
\caption{Effectiveness (nDCG@10) of Qwen2.5-1.5B variants trained after coarse-grained layer dropping across different architectural components. } 
\label{table:qwen2.5-1.5B}
\end{table*}

\begin{table*}[h]
\centering
\scalebox{0.55}{
\begin{tabular}{l|r|rr|rrr}
\toprule
 \textbf{\textit{Qwen2.5-3B}} & Full-Model & \ours-8A & \ours-16A & \ours-8M & \ours-16M & \ours-20M \\
 \midrule
Arguana       &  53.6 & 46.5 & 38.3 & 51.1 & 45.6 & 43.7\\
Climate-FEVER &  25.2 & 20.9 & 19.3 & 24.3 & 20.3 & 21.4\\
DBPedia       &  42.5 & 37.2 & 31.2 & 40.5 & 38.5 & 37.5\\ 
FEVER         &  82.2 & 74.0 & 71.7 & 78.5 & 77.2 & 75.0\\ 
FiQA          &  41.8 & 37.4 & 26.6 & 40.5 & 36.8 & 35.8\\
HotpotQA      &  66.8 & 56.9 & 43.8 & 63.8 & 61.6 & 59.9\\
NFCorpus      &  36.5 & 35.6 & 32.9 & 35.7 & 33.7 & 32.5\\ 
NQ            &  61.1 & 58.3 & 50.4 & 59.1 & 54.0 & 52.7\\ 
Quora         &  88.2 & 86.5 & 83.6 & 87.1 & 85.0 & 84.5\\
SCIDOCS       &  16.5 & 14.5 & 11.1 & 16.8 & 16.1 & 15.8\\
SciFact       &  74.0 & 70.3 & 57.1 & 74.5 & 70.8 & 69.7\\ 
TREC-COVID    &  84.4 & 81.9 & 75.1 & 85.5 & 82.5 & 82.6\\ 
Touche-2020   &  34.0 & 28.7 & 29.5 & 32.9 & 31.5 & 30.3\\ 
\midrule  
Average       & 54.4 & 49.9 & 43.9 & 53.1 & 50.3 & 49.3\\ 
\bottomrule
\end{tabular}
}
\vspace{0.2cm}
\caption{Effectiveness (nDCG@10) of Qwen2.5-3B variants trained after coarse-grained layer dropping across different architectural components. } 
\label{table:qwen2.5-3B}
\end{table*}

\begin{table*}[h]
\centering
\scalebox{0.55}{
\begin{tabular}{l|r|rr|rrr}
\toprule
 \textbf{\textit{Qwen2.5-7B}} & Full-Model & \ours-8A & \ours-16A & \ours-8M & \ours-16M & \ours-20M \\
 \midrule
Arguana       & 53.1&50.0&43.6&51.9&48.5&45.3 \\
Climate-FEVER & 24.2&25.0&22.8&26.7&25.6&21.4 \\
DBPedia       & 44.2&44.3&42.0&44.0&42.3&36.1 \\ 
FEVER         & 81.8&70.3&78.2&82.3&81.2&76.4\\ 
FiQA          & 43.9&37.7&36.6&40.0&38.4&31.3 \\
HotpotQA      & 67.9&65.5&64.3&67.8&65.6&60.0\\
NFCorpus      & 37.1&35.7&34.6&37.4&36.5&32.4\\ 
NQ            & 62.8&60.1&58.2&61.6&57.9&51.1\\ 
Quora         & 88.6&87.7&86.8&87.6&87.2&78.7 \\
SCIDOCS       &  18.1&17.1&15.8&17.8&18.1&15.5\\
SciFact       & 74.1&71.9&69.9&74.0&73.2&69.4 \\ 
TREC-COVID    & 83.7&81.5&86.2&83.8&84.0&77.8\\ 
Touche-2020   & 32.7&25.2&28.5&32.9&31.2&33.2\\ 
\midrule  
Average       & 54.8&51.7&51.3&54.4&53.1&48.4\\ 
\bottomrule
\end{tabular}
}
\vspace{0.2cm}
\caption{Effectiveness (nDCG@10) of Qwen2.5-7B variants trained after coarse-grained layer dropping across different architectural components. } 
\label{table:qwen2.5-7B}
\end{table*}

\begin{table*}[h]
\centering
\scalebox{0.55}{
\begin{tabular}{l|r|rr|rrr}
\toprule
 \textbf{\textit{ModernBERT-base}} & Full-Model & \ours-8A & \ours-16A & \ours-8M & \ours-16M & \ours-20M \\
 \midrule
Arguana       & 51.6&44.7&0.1&41.9&37.0&35.6 \\
Climate-FEVER & 20.1&8.7&0.0&17.6&19.3&12.5 \\
DBPedia       &26.0&12.0&0.0&17.9&14.4&7.6 \\ 
FEVER         &66.4&52.0&0.0&52.3&52.4&32.6\\ 
FiQA          &28.5&11.5&0.2&21.4&17.5&15.1 \\
HotpotQA      &47.7&23.7&0.0&44.4&36.0&32.4\\
NFCorpus      &24.5&22.3&0.6&21.5&19.1&15.9\\ 
NQ            &44.2&35.7&0.0&35.5&27.9&23.4\\ 
Quora         &79.9&69.6&0.0&83.7&79.3&77.0 \\
SCIDOCS       &12.4&2.3&0.0&9.6&7.8&4.7\\
SciFact       &58.6&48.0&0.0&54.6&48.8&40.1 \\ 
TREC-COVID    &77.8&67.1&0.0&72.5&65.4&60.7\\ 
Touche-2020   &26.7&25.8&0.0&24.7&23.4&25.1\\ 
\midrule  
Average       &43.4&32.6&0.1&38.3&34.5&29.4\\ 
\bottomrule
\end{tabular}
}
\vspace{0.2cm}
\caption{Effectiveness (nDCG@10) of ModernBERT-base variants trained after coarse-grained layer dropping across different architectural components. } 
\label{table:modernbert-base}
\end{table*}

\begin{figure*}[h]
  \centering
  \begin{subfigure}[t]{0.31\linewidth}
    \centering
    \includegraphics[width=\linewidth]{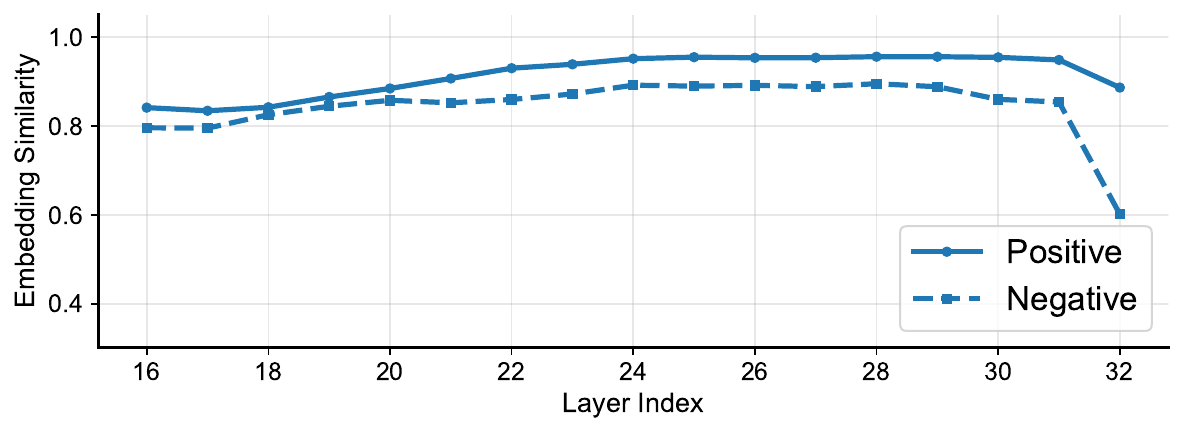}
    \caption{Baseline}
  \end{subfigure}
  \hfill
  \begin{subfigure}[t]{0.31\linewidth}
    \centering
    \includegraphics[width=\linewidth]{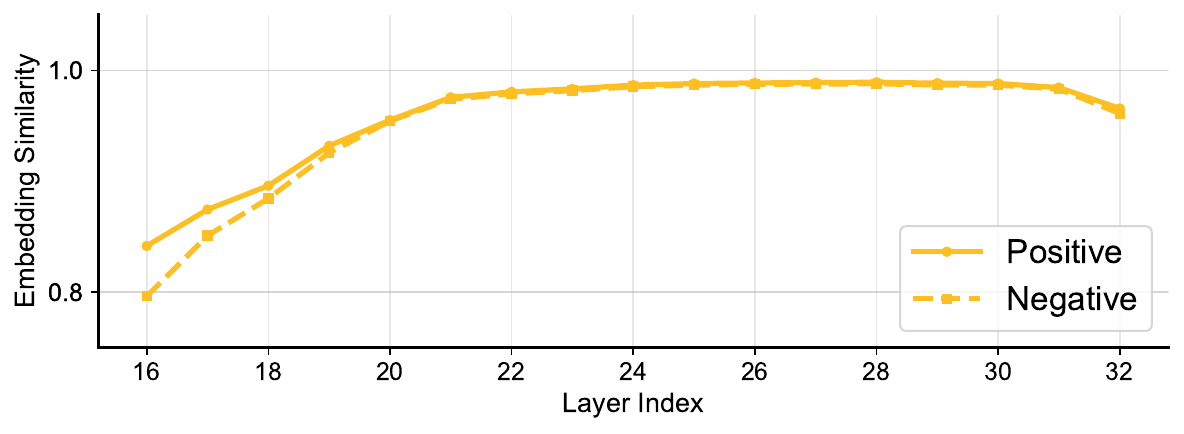}
    \caption{Attention Drop}
  \end{subfigure}
  \hfill
  \begin{subfigure}[t]{0.31\linewidth}
    \centering
    \includegraphics[width=\linewidth]{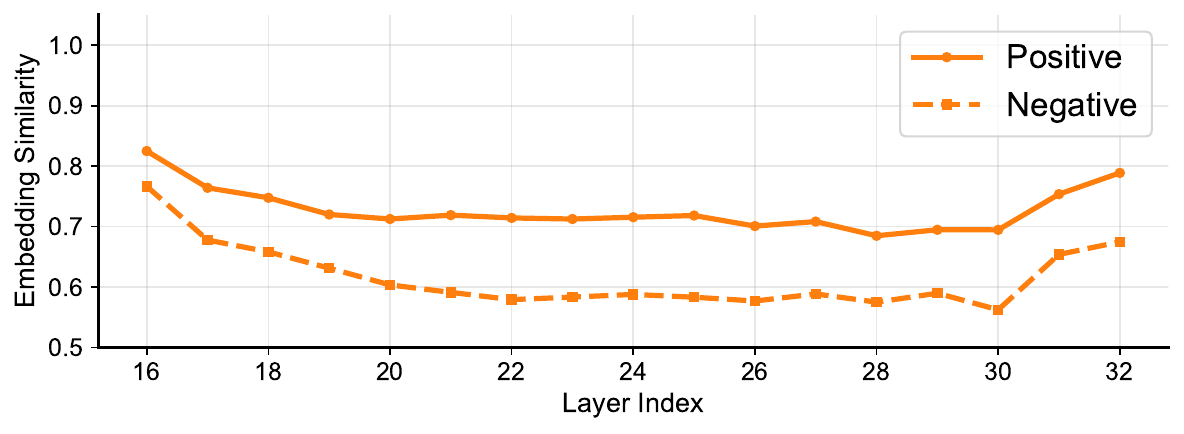}
    \caption{MLP Drop}
  \end{subfigure}

\caption{Layer-wise query–document similarity for last-token-based retrieval in the original model (a) and compressed models (b) and (c). The same query is paired with different documents, including a positive and a negative example. }
    \label{fig:layer-wise_sim_last}
\end{figure*}

As shown in Figure~\ref{fig:layer-wise_sim_last}, models based on last-token embeddings (e.g., E5-Mistral \citep{wang2024improving-text}) lose the ability to distinguish between positive and negative documents when attention layers are dropped, whereas those with reduced MLP layers largely retain this capability. We hypothesize this is because removing attention layers disrupts the aggregation of contextual information, leading to degraded sequence representations. In contrast, pruning MLP layers preserves the information flow from context tokens to the final token, thereby maintaining retrieval effectiveness. 

\begin{figure*}[h]
  \centering
  \begin{subfigure}[t]{0.31\linewidth}
    \centering
    \includegraphics[width=\linewidth]{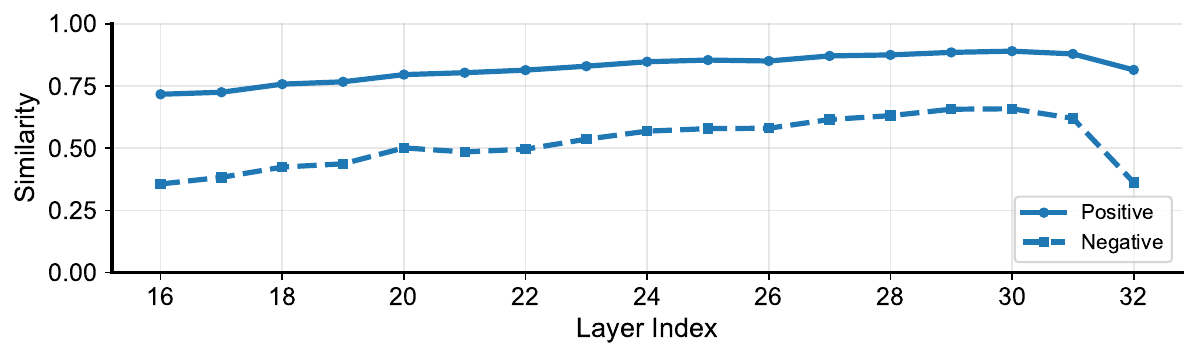}
    \caption{Baseline}
  \end{subfigure}
  \hfill
  \begin{subfigure}[t]{0.31\linewidth}
    \centering
    \includegraphics[width=\linewidth]{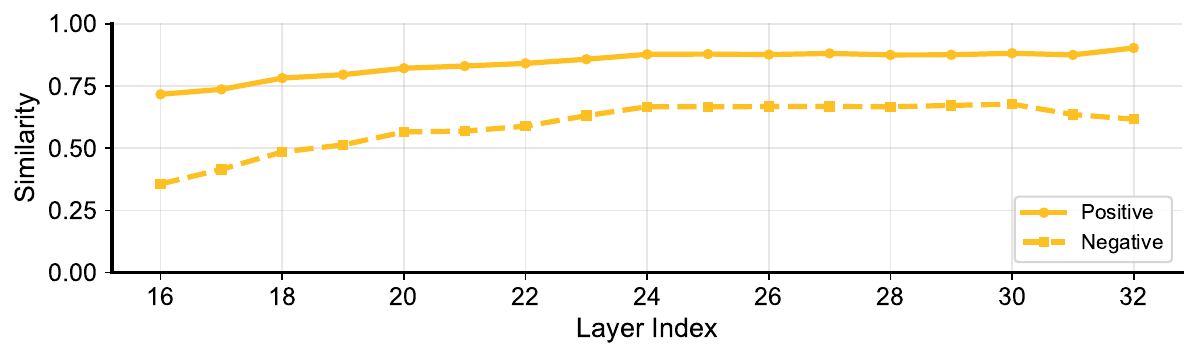}
    \caption{Attention Drop}
  \end{subfigure}
  \hfill
  \begin{subfigure}[t]{0.31\linewidth}
    \centering
    \includegraphics[width=\linewidth]{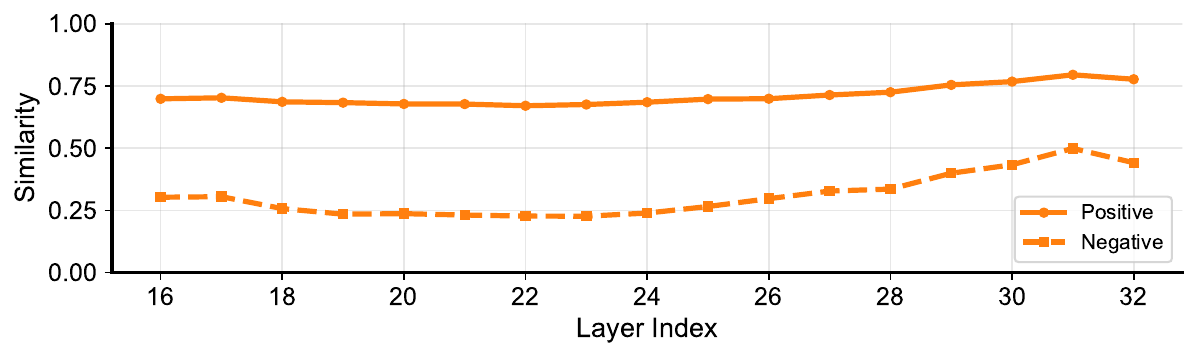}
    \caption{MLP Drop}
  \end{subfigure}

\caption{Layer-wise query–document similarity for mean-pooling-based retrieval in the original model (a) and compressed models (b) and (c). The same query is paired with different documents, including a positive and a negative example. }
    \label{fig:layer-wise_sim_mean}
\end{figure*}

On the other hand, as shown in Figure~\ref{fig:layer-wise_sim_mean}, models using mean-pooling embeddings demonstrate greater robustness to the removal of attention layers compared to those using last-token-based embeddings. This is likely because the mean-pooled embedding still aggregates information from all tokens in the sequence, effectively preserving contextual information.

\end{document}